\documentclass[aps,prb,twocolumn]{revtex4}
\usepackage{amsmath} 
\usepackage{amsfonts} 
\usepackage{listings}
\usepackage{graphicx,stix}

\begin{document}
\title{Introducing SU(3) color charge in undergraduate quantum mechanics}

\author{Brandon L. Inscoe}
\altaffiliation{Present address: Joint School of Nanoscience \& Nanoengineering, 2907 E Gate City Boulevard, Greensboro, NC 27401 USA}
\affiliation{Department of Physics, High Point University, One University Parkway, High Point, NC 27268 USA} %

\author{Jarrett L. Lancaster}
\email[Electronic mail: ]{jlancas2@highpoint.edu}
\affiliation{Department of Physics, High Point University, One University Parkway, High Point, NC 27268 USA} %

\begin{abstract}
We present a framework for investigating effective dynamics of SU(3) color charge. Two- and three-body effective interaction terms inspired by the Heisenberg spin model are considered. In particular, a toy model for a three-source ``baryon'' is constructed and investigated analytically and numerically for various choices of interactions. VPython is used to visualize the nontrivial color charge dynamics. The treatment should be accessible to undergraduate students who have taken a first course in quantum mechanics, and suggestions for independent student projects are proposed.
\end{abstract}
\date{\today}
\maketitle
\section{Introduction}
It is quite notable that one can earn a college degree in physics without significant exposure to two of the four known fundamental interactions. Aspects of gravity and electromagnetism are covered in any introductory sequence, and at least one semester is typically devoted to an in-depth study of electromagnetism. While an investigation of gravity as a field theory is typically reserved for a course on general relativity, it is striking that the strong and weak nuclear interactions are rarely mentioned in any meaningful way until one gets to a graduate-level course in quantum field theory.  

However, it is possible to introduce the relevant structure of nontrivial gauge theories by leaning heavily on the framework of spin, which typically constitutes a significant portion of the standard undergraduate experience in quantum mechanics. Operationally, the quantum mechanics of the spin degree of freedom is deeply connected to the structure of the group $\mbox{SU}(2)$. In the currently-accepted theory of the strong nuclear interaction, quantum {\it chromodynamics},\cite{QCD} the relevant gauge group is $\mbox{SU}(3)$. The underlying field theory describing the strong interaction bares a formal similarity to that of electromagnetism (i.e., Maxwell's equations) but with a much richer structure due to the nonabelian\cite{note1} gauge group. Rather than attempt to introduce quantum chromodynamics as a fully-formed field theory, the present aim is to explore the structure imposed by $\mbox{SU}(3)$ in the context of quantum mechanics. To this end, we present a quantum mechanical system consisting of sources of $\mbox{SU}(3)$ ``color'' charge and explore the dynamics generated by effective interactions. In addition to having pedagogical value, this model may be relevant for simulations involving cold atoms in optical lattices.

This paper is organized as follows. Some basic aspects of quantum mechanical spin are summarized in Sec.~\ref{sec:spin}. While much of this material is fairly standard, our treatment of $\mbox{SU}(3)$ dynamics follows the setup presented for spin rather closely. Section~\ref{sec:color} introduces the notion of ``color charge'' as a generalization of electric charge which shares similarities with the structure of spin. By analogy with the magnetic interaction of two spins, effective interactions for color charges with external fields and other color charges are proposed. In Sec.~\ref{sec:dynamics} we explore the color charge dynamics in systems of interacting color sources with particular attention paid to a three-body system which serves as a toy model for a baryon. Lastly, conclusions and suggestions for student projects are contained in Sec.~\ref{sec:conclusion}

\section{Spin}
\label{sec:spin}
Before defining color charge, it is useful to review some basic aspects of the quantum mechanics of spin. Here we present some introductory results from which our treatment of $\mbox{SU}(3)$ sources and interactions will follow by analogy. Specifically, we require the basic properties of several, interacting, spin-$\frac{1}{2}$ degrees of freedom.

\subsection{One spin}\label{sec:onespin}
A single spin-$\frac{1}{2}$ degree of freedom is described by a state $\left|\chi\right\rangle$ which belongs to the fundamental representation of $\mbox{SU}(2)$ and can be represented by a two-component column vector
\begin{eqnarray}
\left|\chi\right\rangle & = & a\left|\uparrow\right\rangle + b\left|\downarrow\right\rangle \dot{=}  \left(\begin{array}{c} a\\ b\end{array}\right),\label{eq:spinstate}
\end{eqnarray}
for some complex numbers $a$ and $b$ satisfying $|a|^{2} + |b|^{2} = 1$. Here the symbol $\dot{=}$ stands for ``represented by'' in the sense that the abstract, two-dimensional quantum state may be represented by a two-component, column vector. Operators acting on these states will be represented by $2\times 2$ matrices, and the action of operators on states is represented by ordinary matrix multiplication.

Given a state $\left|\chi\right\rangle$, the expectation value of some observable, such as one of the three components of spin, can be computed as the inner product $\left\langle \hat{S}^{x}\right\rangle = \left\langle \chi\right|\hat{S}^{x}\left|\chi\right\rangle$. The spin operators are $\hat{S}^{\nu} \dot{=} \frac{\hbar}{2}\sigma^{\nu}$, where $\nu = x,y,z$ and the Pauli matrices $\sigma^{\nu}$ given by
\begin{eqnarray}
\sigma^{x} & = & \left(\begin{array}{cc} 0 & 1 \\ 1 & 0\end{array}\right),\;\;\;\;
\sigma^{y} = \left(\begin{array}{cc} 0 & -i \\ i & 0\end{array}\right),\;\;\;\;
\sigma^{z} = \left(\begin{array}{cc} 1 & 0 \\ 0 & -1\end{array}\right).\label{eq:pauli}
\end{eqnarray}
We note that in this representation the squared total spin operator takes the form
\begin{eqnarray}
\hat{S}^{2} & = & \left[\hat{S}^{x}\right]^{2} +\left[\hat{S}^{y}\right]^{2}  + \left[\hat{S}^{z}\right]^{2} \dot{=} \frac{3\hbar^{2}}{4}\hat{I},\label{eq:s2single}
\end{eqnarray}
where $\hat{I}$ is the identity matrix. That is, $\hat{S}^{2}$ commutes with each component of spin. A general spin-$s$ system can always be decomposed into a basis in which each basis state has a well-defined value of total spin $s$ and $z$-component of spin $s_{z}$,
\begin{eqnarray}
\hat{S}^{2}\left|s,s_{z}\right\rangle & = & s(s+1)\hbar^{2}\left|s,s_{z}\right\rangle,\nonumber\\
\hat{S}^{z}\left|s,s_{z}\right\rangle & = & s_{z}\hbar\left|s,s_{z}\right\rangle.\label{eq:szsingle}
\end{eqnarray}
For a single spin, the abbreviated notation $\left|\uparrow\right\rangle \rightarrow \left|\frac{1}{2},\frac{1}{2}\right\rangle$, $\left|\downarrow\right\rangle\rightarrow \left|\frac{1}{2},-\frac{1}{2}\right\rangle$ is convenient when $s = \frac{1}{2}$ is clear from context. 

To generate nontrivial dynamics for a single degree of freedom, the spin must interact with other spins or external magnetic fields. Postponing discussion of multiple sources to the next subsection, the Zeeman interaction of a single spin with an external magnetic field is encoded in the Hamiltonian $\hat{H}_{1}$,
\begin{eqnarray}
\hat{H}_{1} & = & -\lambda {\bf B}\cdot\hat{\bf S},\label{eq:ham1}
\end{eqnarray}
where $\lambda$ is a constant proportional to the magnetic moment of the source, and ${\bf B} = B_{x}{\bf \hat{x}} + B_{y}{\bf \hat{y}}  + B_{z}{\bf \hat{z}} $ is an applied magnetic field. We employ the shorthand $\hat{\bf S} = \hat{S}^{x}{\bf \hat{x}} +\hat{S}^{y}{\bf \hat{y}} + \hat{S}^{z}{\bf \hat{z}}$ where each coefficient $\hat{S}^{\nu}$ is actually an operator represented by a $2\times 2$ matrix equal to $\frac{\hbar}{2}$ times a Pauli matrix in Eq.~(\ref{eq:pauli}). Thus, the Hamiltonian Eq.~(\ref{eq:ham1}) takes the form ,
\begin{eqnarray}
\hat{H}_{1} & \dot{=} & -\frac{\lambda \hbar}{2}\left(\begin{array}{cc} B_{z} & (B_{x}-iB_{y})\\ (B_{x}+iB_{y}) & - B_{z}\end{array}\right).
\end{eqnarray}
As a simple example, let us take ${\bf B}$ to be a constant. The fundamental problem in quantum mechanics is to obtain the time-dependent expectation value of observables $O(t) = \left\langle \chi(t)\right|\hat{O}\left|\chi(t)\right\rangle$ for some initial state $\left|\chi(0)\right\rangle \equiv \left|\chi_{0}\right\rangle$ and operator $\hat{O}$. For a general state given by Eq.~(\ref{eq:spinstate}), the spin expectation values can be computed by using the Pauli matrices in Eq.~(\ref{eq:pauli}),
\begin{eqnarray}
\left\langle \chi(t) \right| \hat{S}^{x} \left|\chi(t)\right\rangle & = & \frac{1}{2}\left(a^{*}(t)b(t) + b^{*}(t)a(t)\right),\label{eq:spintime1}\\
\left\langle \chi(t) \right| \hat{S}^{y} \left|\chi(t)\right\rangle & = & \frac{1}{2i}\left(a^{*}(t)b(t) - b^{*}(t)a(t)\right),\label{eq:spintime2}\\
\left\langle \chi(t) \right| \hat{S}^{z} \left|\chi(t)\right\rangle & = & \frac{1}{2}\left(|a(t)|^{2} - |b(t)|^{2}\right).\label{eq:spintime3}
\end{eqnarray}
Explicit forms for the complex amplitudes $a(t)$ and $b(t)$ are obtained at arbitrary times by solving the time-dependent Schr\"{o}dinger equation
\begin{eqnarray}
i\hbar\frac{\partial \left|\chi(t)\right\rangle}{\partial t} & =& \hat{H}\left|\chi(t)\right\rangle.\label{eq:schrod1}
\end{eqnarray}
In the present situation, Eq.~(\ref{eq:schrod1}) reduces to a system of two coupled, linear, differential equations for $a(t)$ and $b(t)$,
\begin{eqnarray}
i\dot{a}(t) & = & -\frac{\lambda}{2}\left[B_{z}a(t) + (B_{x}-iB_{y})b(t)\right],\nonumber\\
i\dot{b}(t) & = & -\frac{\lambda}{2}\left[(B_{x}+iB_{y})a(t) - B_{z}b(t)\right].\label{eq:spindynamics1}
\end{eqnarray}
For concreteness, let us take $B_{y}=B_{0}$, $B_{x}=B_{z}=0$ and consider an initial state given by Eq.~(\ref{eq:spinstate}) with $a(0) = 1$ and $b(0) = 0$. Then Eq.~(\ref{eq:spindynamics1}) gives
\begin{eqnarray}
\left|\chi(t)\right\rangle & \dot{=} & \left(\begin{array}{c} \cos(\omega t)\\ -\sin(\omega t)\end{array}\right),
\end{eqnarray}
with $\omega = \frac{\lambda B_{0}}{2}$. Using Eq.~(\ref{eq:schrod1}), one may compute the time-dependent expectation value of each spin component,
\begin{eqnarray}
\left\langle \hat{S}^{x}(t)\right\rangle & = & -\frac{\hbar}{2}\sin[2\omega t],\nonumber\\
\left\langle \hat{S}^{y}(t)\right\rangle & = &0,\nonumber\\
\left\langle \hat{S}^{z}(t)\right\rangle & = & \frac{\hbar}{2}\cos[2\omega t].\label{eq:spinprecess}
\end{eqnarray}
Thus, the spin precesses about the external magnetic field. While this is not a particularly complicated example, the steps involved in obtaining Eqs.~(\ref{eq:spinprecess}) are virtually identical to those we will take with $\mbox{SU}(3)$ sources in later sections.
\subsection{Multiple spins}
A single spin interacts only with external magnetic fields which couple to its magnetic moment. When brought into the vicinity of another spin, the two spins may also interact with each other.\cite{note3} Before addressing interactions, we summarize the formalism for representing a system of two spins. Given two spins in quantum states $\left|\chi_{1}\right\rangle$ and $\left|\chi_{2}\right\rangle$, the total quantum state is constructed by forming a tensor product of the individual spin states,
\begin{eqnarray}
\left|\chi\right\rangle & = & \left|\chi_{1}\right\rangle \otimes \left|\chi_{2}\right\rangle.
\end{eqnarray}
It is convenient to represent the tensor product as a Kronecker product so that
\begin{eqnarray}
\left|\chi_{1}\right\rangle \otimes \left|\chi_{2}\right\rangle & \dot{=} & \left(\begin{array}{c} a_{1}\\ b_{1}\end{array}\right)\otimes\left(\begin{array}{c} a_{2}\\ b_{2}\end{array}\right)
\equiv \left(\begin{array}{c} a_{1}a_{2} \\ a_{1}b_{2}\\ b_{1}a_{2}\\ b_{1}b_{2}\end{array}\right).\label{eq:kronecker}
\end{eqnarray}
Operators acting on the full state are built from single-spin operators and must have the same overall dimension as the total Hilbert space. For example, the expectation value of the $x$-component of the first particle's spin can be obtained from the inner product
\begin{eqnarray}
S_{1}^{x} & = & \left\langle \chi\right| \hat{S}^{x}\otimes \hat{I}\left|\chi\right\rangle,
\end{eqnarray}
where $\hat{I}$ is the identity operator, represented in this context by the $2\times 2$ identity matrix. The total $x$ component of spin for the two-spin system is represented by
\begin{eqnarray}
\hat{S}^{x} & = & \hat{S}_{1}^{x} + \hat{S}_{2}^{x} \equiv \hat{S}^{x}\otimes\hat{I} + \hat{I}\otimes \hat{S}^{x}.
\end{eqnarray}

The simplest phenomenological interaction for two spins is known as the Heisenberg Hamiltonian,\cite{Santos} which takes the form $\hat{H}_{2} = -J\hat{\bf S}_{1}\cdot\hat{\bf S}_{2}$, where the dot product is shorthand for
\begin{eqnarray}
\hat{H}_{2} & = & -J\left[\hat{S}^{x}\otimes \hat{S}^{x} + \hat{S}^{y}\otimes \hat{S}^{y} + \hat{S}^{z}\otimes \hat{S}^{z}\right],\label{eq:heisenberg}
\end{eqnarray}
Here $J$ is an effective interaction energy, and the explicit representation of $\hat{H}_{2}$ takes the form of a $2^{2}\times 2^{2} = 4\times 4$ matrix. Proceeding in this manner, one can construct spin systems of any size with arbitrary interactions. However, the resulting Hamiltonian grows explosively in size with the dimension being $2^{N}$, representing a practical upper limit on the sizes of quantum spin systems which can be simulated effectively on classical computers. In the present work, focus is restricted to small systems with $N \leq 3$.

\section{Color charge}
\label{sec:color}
Formally, the Pauli matrices in Eqs.~(\ref{eq:pauli}) are the generators of $\mbox{SU}(2)$, or the group of $2\times 2$ unitary matrices with determinant equal to unity.\cite{note3a} The group $\mbox{SU}(3)$ is simply the group of $3\times 3$ unitary matrices with unit determinant. Some familiarity with the Pauli matrices, perhaps gained through a study of spin, provides some valuable intuition for the mechanics of working with objects in $\mbox{SU}(3)$.

The eight\cite{note4} Gell-Mann matrices $\lambda^{(\alpha)}$ are to SU(3) what the Pauli matrices are to SU(2). Just as the actual spin--$\frac{1}{2}$ operators are related to the Pauli matrices by factor of $\frac{1}{2}$, one can define color charge operators $t^{(\alpha)} = \frac{1}{2}\lambda^{(\alpha)}$, as
\begin{eqnarray}
\hat{t}^{(1)} & = & \frac{1}{2}\left(\begin{array}{ccc} 0 & 1 & 0 \\ 1 & 0 & 0\\ 0 & 0 & 0\end{array}\right) \;\;\; \hat{t}^{(2)} = \frac{1}{2}\left(\begin{array}{ccc} 0 & -i & 0 \\ i & 0 & 0\\ 0 & 0 & 0\end{array}\right),\\
\hat{t}^{(3)} & = & \frac{1}{2}\left(\begin{array}{ccc} 1 & 0 & 0 \\ 0 & -1 & 0\\ 0 & 0 & 0\end{array}\right)\;\;\; \hat{t}^{(4)} = \frac{1}{2}\left(\begin{array}{ccc} 0 & 0 & 1 \\ 0 & 0 & 0\\ 1 & 0 & 0\end{array}\right),\\
 \hat{t}^{(5)} & = & \frac{1}{2}\left(\begin{array}{ccc} 0 & 0 & -i \\ 0 & 0 & 0\\ i & 0 & 0\end{array}\right) \;\;\; \hat{t}^{(6)} = \frac{1}{2}\left(\begin{array}{ccc} 0 & 0 & 0 \\ 0 & 0 & 1\\ 0 & 1 & 0\end{array}\right),\\
\hat{t}^{(7)}& = & \frac{1}{2}\left(\begin{array}{ccc} 0 & 0 & 0 \\ 0 & 0 & -i\\ 0 & i & 0\end{array}\right)\;\;\; \hat{t}^{(8)} = \frac{1}{2\sqrt{3}}\left(\begin{array}{ccc} 1 & 0 & 0 \\ 0 & 1 & 0\\ 0 & 0 & -2\end{array}\right).\label{eq:gellmann}
\end{eqnarray}
For brevity, we shall sometimes refer to the $\hat{t}^{(\alpha)}$ as the Gell-Mann matrices. By analogy with spin, a single source of $\mbox{SU}(3)$ ``color charge'' belongs to the fundamental representation\cite{note4a} of $\mbox{SU}(3)$ and is represented by a three-component vector
\begin{eqnarray}
\left|\psi\right\rangle & = & \alpha \left|r\right\rangle +  \beta \left|g\right\rangle +  \gamma \left|b\right\rangle \dot{=}\left(\begin{array}{c} \alpha \\ \beta \\ \gamma\end{array}\right).\label{eq:psi0}
\end{eqnarray}
Here the basis states corresponding to ``up'' and ``down'' in $\mbox{SU}(2)$ correspond to the three basis states ``red,'' ``green'' and ``blue'' in a theory built upon SU(3). The motivation for using colors to label these basis states will be explained below. The treatment of color charge that follows will use minimal group theory, but the interested reader is encouraged to explore the role of group theory in the strong interaction through one of many excellent sources on the subject.\cite{Swart,ZeeGroup} It is useful to label explicitly the three basis vectors which span the color space
\begin{eqnarray}
 \left|r\right\rangle & \dot{=} & \left(\begin{array}{c} 1\\ 0 \\ 0 \end{array}\right),\;\;\;\;
 \left|g\right\rangle  \dot{=} \left(\begin{array}{c} 0\\ 1 \\ 0 \end{array}\right),\;\;\;\;
 \left|b\right\rangle \dot{=} \left(\begin{array}{c} 0\\ 0 \\ 1 \end{array}\right).\label{eq:colorbasis}
 \end{eqnarray}
Just as each Pauli matrix corresponds to a component of spin, each Gell-Mann matrix corresponds to a component of color charge $Q^{(\alpha)}$ with
\begin{eqnarray}
Q^{(\alpha)} & \rightarrow & \left\langle \psi \right|\hat{t}^{(\alpha)}\left|\psi\right\rangle.\label{eq:chargecomp}
\end{eqnarray}
Color charge is a type of generalization of electric charge possessed by quarks. In addition to color charge, quarks also have electric charge, flavor and spin. We will sometimes use the term ``quark'' to refer to point sources of color charge, but we will consider only the color charge degree of freedom. It is important to note that the color basis states of red, green and blue are analogous to spin-up and spin-down, while the eight components of color {\it charge} are analogous to the three components of spin. Thus it is critical to distinguish between ``color state'' and ``color charge.'' Further muddying the waters is the observation that color charge is necessarily an abstract quantity living in eight-dimensional ``gauge space,'' while spin components refer to the components of a three-dimensional vector quantity in {\it physical} space.

However, it is potentially misleading to think of quantum spin as a ``vector,'' since it is impossible to measure all components simultaneously. Mathematically, it is impossible to find a basis in which all three spin operators are represented by diagonal matrices. One is only free to choose a single component---traditionally, the $z$-component---to be diagonal in addition to the squared spin, $\hat{S}^{2}$ as in Eq.~(\ref{eq:szsingle}). In the famous Stern-Gerlach experiment,\cite{McIntyre} the incompatibility of various spin components is shown to lead to entirely {\it un}classical behavior.

\begin{figure}[h]
\begin{center}
\includegraphics[totalheight=13.5cm]{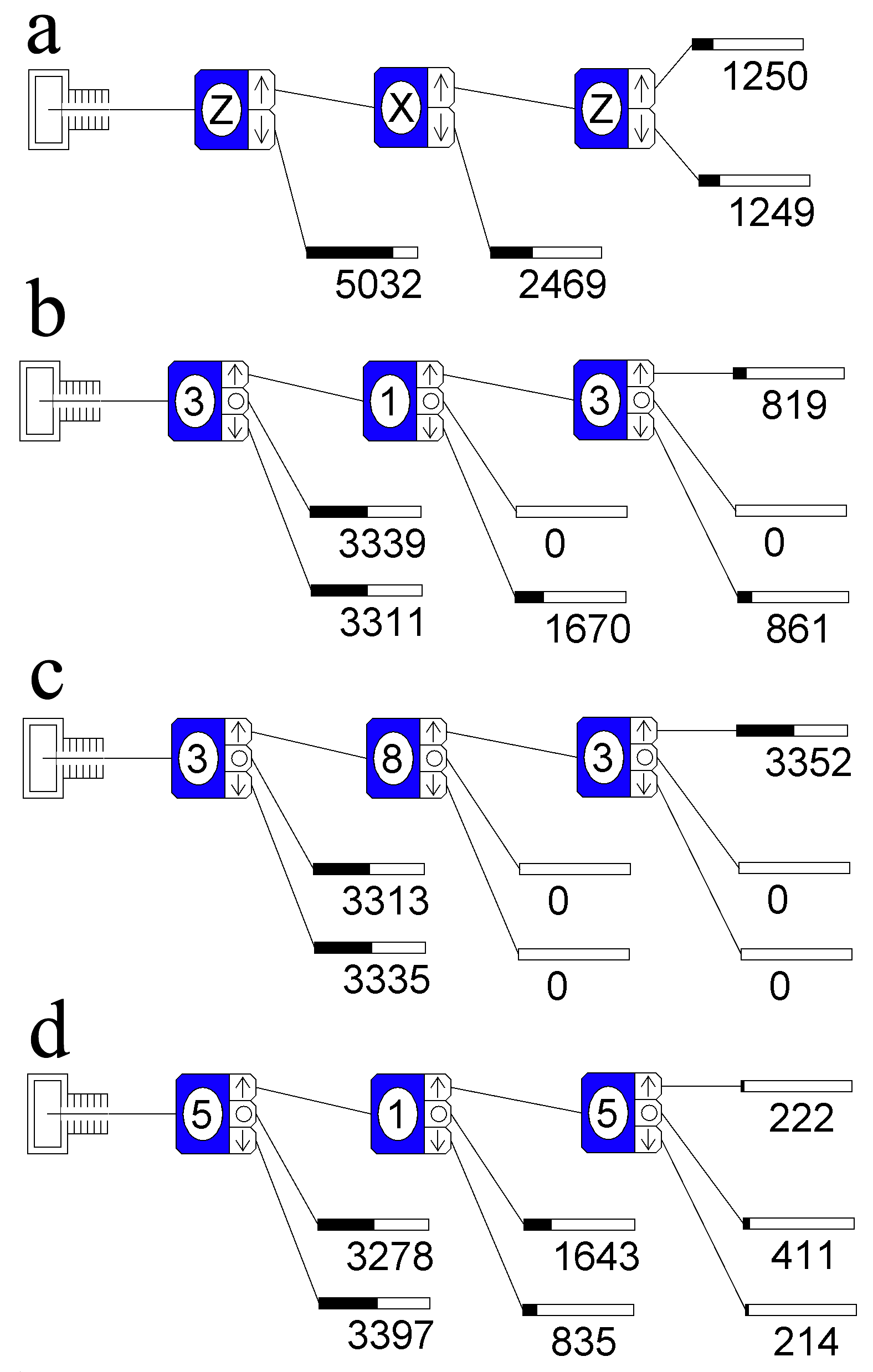}
\caption{Stern--Gerlach experiment simulation using \textsc{Spins} software for (a) spin--$\frac{1}{2}$ particles and (c)--(d) SU(3) color charge. In each case, an unpolarized ``beam'' of 10,000 particles is fed into a particular sequence of analyzers. Since SU(2) is a subgroup of SU(3), results are entirely analogous to spin--$\frac{1}{2}$ when only operators $\hat{t}^{(1)}$, $\hat{t}^{(2)}$, $\hat{t}^{(3)}$ are used, as shown in panel (b). The existence of two commuting operators (namely $\hat{t}^{(3)}$ and $\hat{t}^{(8)}$) allows for both operators to have simultaneously well-defined values as shown in (c). Panel (d) depicts a more complex example of the SU(3) case.}
\label{fig:sg}
\end{center}
\end{figure}

A particularly instructive form of the experiment involves ``analyzing'' a beam of unpolarized spins by physically separating the $\left|\uparrow\right\rangle$ states from the $\left|\downarrow\right\rangle$ states. Upon measuring $\hat{S}^{x}$ on the $\left|\uparrow\right\rangle$ states, those spins which return $+\frac{\hbar}{2}$ are subjected to a second measurement of $\hat{S}^{z}$. It is found that the states split evenly into $\left|\uparrow\right\rangle$ and $\left|\downarrow\right\rangle$ upon this second measurement of $\hat{S}^{z}$. That is, the measurement process affects the state of the system. Simulations of this type of experiment can be performed using the freely-available, \textsc{Java}-based \textsc{Spins} software.\cite{McIntyreSpins,SchroederSpins} A screenshot of a simulation of the case described above with 10,000 spins is depicted in panel (a) of Fig.~\ref{fig:sg}. Here, the measurement of $\hat{S}^{x}$ projects the states $\left|\uparrow\right\rangle$ onto the eigenstates of $\hat{S}^{x}$, $\frac{1}{\sqrt{2}}\left(\left|\uparrow\right\rangle \pm\left| \downarrow\right\rangle\right)$ so that a subsequent measurement of $\hat{S}^{z}$ is equally likely to return $+\frac{\hbar}{2}$ or $-\frac{\hbar}{2}$.

The \textsc{Spins} program also allows one to simulate Stern-Gerlach-like experiments with the spin--$\frac{1}{2}$ particles replaced by spin--$1$ particles or by particles possessing SU(3) color charge. The program's notation for SU(3) states is inspired by the convention for labeling spin--1 states, with the correspondence between the notation in this paper and the program being $\left|r\right\rangle \rightarrow \left|\uparrow\right\rangle$, $\left|b\right\rangle \rightarrow \left|0\right\rangle$, $\left|g\right\rangle \rightarrow \left|\downarrow\right\rangle$. Since the first three generators of SU(3) are essentially the spin--$\frac{1}{2}$ operators which generate SU(2) on the red and green states, any set of analyzers using only $\hat{t}^{(1)}$, $\hat{t}^{(2)}$ and $\hat{t}^{(3)}$ maps exactly to a spin--$\frac{1}{2}$ experiment aside from the blue states which are obtained from the first analyzer. Subsequent analyzers built from these first three generators yield no additional blue states since the SU(2) subgroup is closed within the red and green states, as shown in panel (b) of Fig.~\ref{fig:sg}.  

The indeterminacy of certain pairs of observables results mathematically when the two operators do not commute. In SU(3), both $\hat{t}^{(3)}$ and $\hat{t}^{(8)}$ are represented by diagonal matrices which commute with each other. Consequently, as shown in panel (c) of Fig.~\ref{fig:sg}, the values of these two observables {\it can} be simultaneously well-defined. 

Moving away from the commuting generators and SU(2) subgroup, panel (d) of Fig.~\ref{fig:sg} depicts one of the more complex analyzer constructions which ``mixes'' all three color states. The action of the first analyzer corresponds to a measurement of the color charge operator $\hat{t}^{(5)}$. The reader may verify that the eigenvalues of $\hat{t}^{(5)}$ are $0,\pm \frac{1}{2}$ with eigenstates given by
\begin{eqnarray}
\left|+\right\rangle_{5} & = & \frac{1}{\sqrt{2}}\left(\left|r\right\rangle +i\left|b\right\rangle\right),\nonumber\\
\left|0\right\rangle_{5} & = & \left|g\right\rangle ,\nonumber\\
\left|-\right\rangle_{5} & = & \frac{1}{\sqrt{2}}\left(\left|r\right\rangle -i\left|b\right\rangle\right).
\end{eqnarray}
The first analyzer yields the state $\left|+\right\rangle_{5}$, which is then fed into a measurement of $\hat{t}^{(1)}$, whose eigenvalues are also $0,\pm\frac{1}{2}$. According to the basic postulates of quantum mechanics,\cite{McIntyre} the probability of this second measurement returning the value $v$ is $\left|_{5}\left\langle + \right.\left| v\right\rangle_{1}\right|^{2}$, where $\left|v\right\rangle_{1}$ is the eigenstate of $\hat{t}^{(1)}$ corresponding to eigenvalue $v$. By obtaining the eigenstates of $\hat{t}^{(1)}$ and computing the various inner products, it is straightforward to verify that the simulation yields results consistent with theory. In particular, one may show that the probability $p$ of a particle making it to the final analyzer and being in each of the three possible final states is
\begin{eqnarray}
p\left(\left|+\right\rangle_{5}\right) & = & p\left(\left|-\right\rangle_{5}\right) = \frac{1}{48}, \;\;\;\;\; p\left(\left|0\right\rangle_{5}\right) = \frac{1}{24}.\label{eq:probsg}
\end{eqnarray}
Note that in this case, the probabilities do {\it not} add to one because Eq.~(\ref{eq:probsg}) only describes the probabilities corresponding to a particle making it through all three analyzers.

\begin{figure}[h]
\begin{center}
\includegraphics[totalheight=8.75cm]{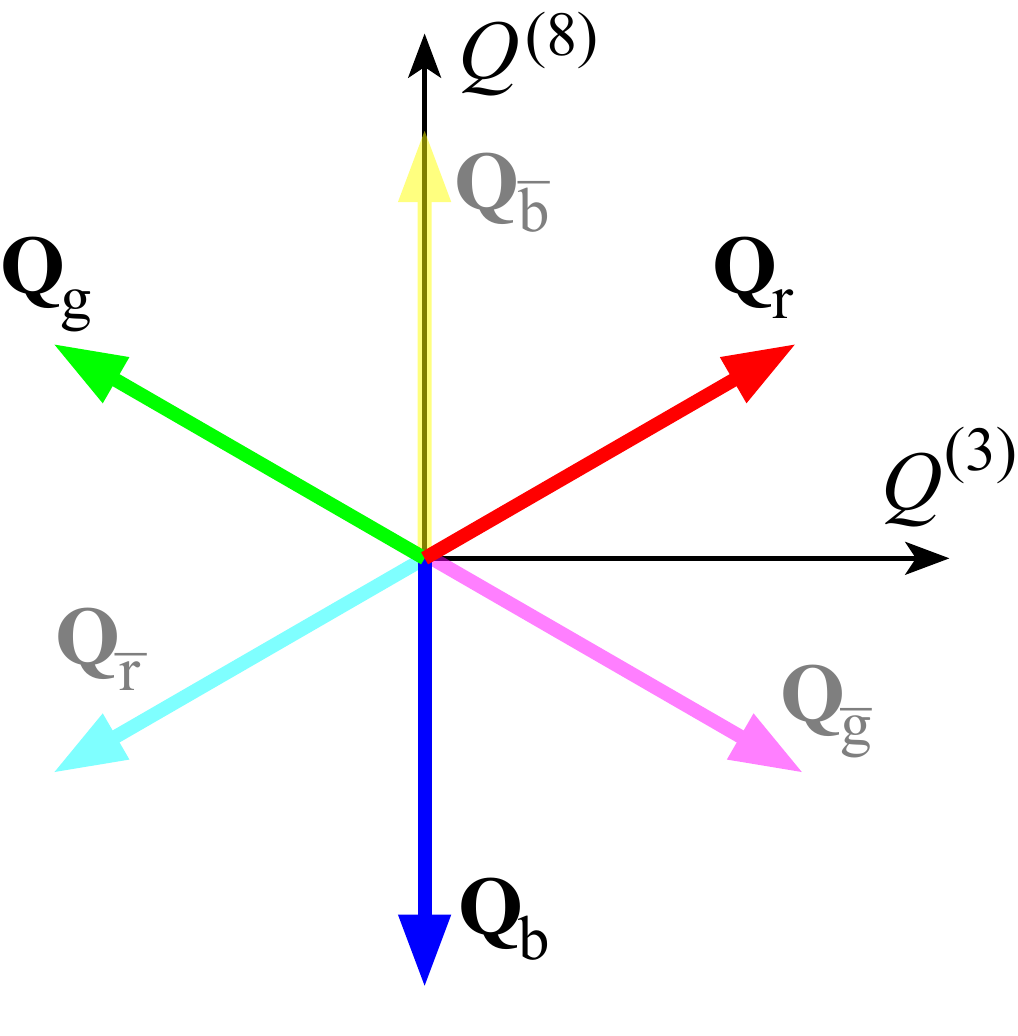}
\caption{Graphic depiction of the vectors ${\bf Q}_{\mbox{\scriptsize r}}$, ${\bf Q}_{\mbox{\scriptsize g}}$, ${\bf Q}_{\mbox{\scriptsize b}}$ and corresponding ``anticolors,'' ${\bf Q}_{\overline{\mbox{\scriptsize r}}}$, ${\bf Q}_{\overline{\mbox{\scriptsize g}}}$ and ${\bf Q}_{\overline{\mbox{\scriptsize b}}}$. Only the $Q^{(3)}$ and $Q^{(8)}$ components are nonzero.}
\label{fig:colorspace}
\end{center}
\end{figure}

Using Eq.~(\ref{eq:chargecomp}), one may show that at most only two components of color charge are nonzero for each of the three basis states in Eq.~(\ref{eq:colorbasis}). Denoting 
\begin{eqnarray}
{\bf Q}_{\mbox{\scriptsize c}} & = & \sum_{\alpha = 1}^{8}Q_{\mbox{\scriptsize c}}^{(\alpha)}\hat{\bf e}^{(\alpha)},\label{eq:cvec}
\end{eqnarray}
where  $Q^{(\alpha)}_{\mbox{\scriptsize c}} = \left\langle \mbox{c}\right| \hat{t}^{(\alpha)}\left|\mbox{c}\right\rangle$ for $\mbox{c} = \mbox{r,g,b}$, and $\hat{\bf e}^{(\alpha)}$ is a unit vector in the $\alpha$ direction in eight-dimensional gauge space. An explicit representation for these abstract, basis vectors is provided by the basis
\begin{eqnarray}
\hat{\bf e}^{(1)} & \dot{=} & \left(\begin{array}{c}1\\ 0\\0 \\0 \\0\\0 \\0 \\0\end{array}\right),\;\;\; \hat{\bf e}^{(2)} \dot{=} \left(\begin{array}{c}0\\ 1\\0 \\0 \\0\\0 \\0 \\0\end{array}\right),\;\;\; \hat{\bf e}^{(3)} \dot{=} \left(\begin{array}{c}0\\ 0\\1 \\0 \\0\\0 \\0 \\0\end{array}\right),\cdots.
\end{eqnarray}
Using Eq.~(\ref{eq:cvec}), one may obtain the expectation values of the charge components. For example, a red source has
\begin{eqnarray}
Q_{\mbox{\scriptsize r}}^{(3)} & = & \left(\begin{array}{ccc} 1 & 0 & 0 \end{array}\right)\frac{1}{2}\left(\begin{array}{ccc} 1 & 0 & 0 \\ 0 & -1 & 0 \\ 0 & 0 & 0\end{array}\right)\left(\begin{array}{c} 1 \\ 0 \\ 0 \end{array}\right)= \frac{1}{2}.
\end{eqnarray}
Performing this procedure for all eight components of all three color states using Eqs.~(\ref{eq:gellmann}), (\ref{eq:colorbasis}) yields
\begin{eqnarray}
{\bf Q}_{\mbox{\scriptsize r}} & = & \frac{1}{2}\hat{\bf e}^{(3)} + \frac{1}{2\sqrt{3}}\hat{\bf e}^{(8)},\label{eq:rgbcharge1}\\
{\bf Q}_{\mbox{\scriptsize g}} & = & -\frac{1}{2}\hat{\bf e}^{(3)} + \frac{1}{2\sqrt{3}}\hat{\bf e}^{(8)},\label{eq:rgbcharge2}\\
{\bf Q}_{\mbox{\scriptsize b}} & = & - \frac{1}{\sqrt{3}}\hat{\bf e}^{(8)},\label{eq:rgbcharge3}
\end{eqnarray}
The vanishing of all but the third and eighth components of charge in Eqs.~(\ref{eq:rgbcharge1})--(\ref{eq:rgbcharge3}) is analogous to the $x$- and $y$-components of spin expectation values being zero for the basis states $\left|\uparrow\right\rangle$ and $\left|\downarrow\right\rangle$. While $\sigma^{z}$ is the only diagonal Pauli matrix (corresponding to the only nonzero component of spin in the $z$-basis), there are two diagonal Gell-Mann matrices, $\hat{t}^{(3)}$ and $\hat{t}^{(8)}$. As diagonal matrices, the corresponding color charge operators will commute and can thus be observed simultaneously, as demonstrated in the simulation shown in Fig.~\ref{fig:sg} (c). That is, $Q^{(3)}$ and $Q^{(8)}$ correspond to well-defined quantum numbers. A plot of the color charge vectors ${\bf Q}_{\mbox{\scriptsize r,g,b}}$ is shown in Fig.~\ref{fig:colorspace}. Since most charge components have vanishing expectation values, it is convenient to project the abstract vectors onto the $(Q^{(3)},Q^{(8)})$ plane. That is, for the purpose of visual depiction, we will adopt the representation $\hat{\bf e}^{(3)} \dot{=} \left(\begin{array}{c} 1\\ 0\end{array}\right)$, $\hat{\bf e}^{(8)} \dot{=} \left(\begin{array}{c} 0\\ 1\end{array}\right)$ since all components not shown are zero unless otherwise stated.

It should be stressed that there is no physical significance to the use of color in labeling the three basis states. The ``colorful'' nomenclature is inspired by similarity between the vector nature of these charges and that of adding different colors of light. Just as red, blue and green light combine to form white (colorless) light, these three vectors add to zero. Additionally, to each quark there is also an antiquark containing the ``opposite'' charge. While the electron has electric charge $-e$, the positron carries charge $-(-e) = +e$. The same basic reasoning applies to color charge so that an ``anti-red'' quark carries color charge ${\bf Q}_{ \overline{\mbox{\scriptsize r}}} = -{\bf Q}_{\mbox{\scriptsize r}}$. 

One may gain some intuition for the relationship between the three states and eight charge operators by considering the action of changing one basis state into another. For example, let us consider the operator $\hat{M}_{r\rightarrow g}$ which changes a red quark into a green quark,
\begin{eqnarray}
\hat{M}_{r\rightarrow g}\left|r\right\rangle & = & \left|g\right\rangle.
\end{eqnarray}
Making use of the tensor product employed in the previous section, one can form the matrix representation of $\hat{M}_{r\rightarrow g}$ as the outer product,
\begin{eqnarray}
\hat{M}_{r \rightarrow g} & \equiv & \left| g\right\rangle \left\langle r\right|  \dot{=} \left(\begin{array}{c} 0 \\ 1\\ 0\end{array}\right)\left(\begin{array}{ccc} 1 & 0 & 0\end{array}\right),\nonumber\\
& = & \left(\begin{array}{ccc} 0 & 0 & 0 \\ 1 & 0 & 0 \\ 0 & 0 & 0\end{array}\right).
\end{eqnarray}
where the mathematical representation of this outer product takes a form quite similar to that of the tensor product introduced in the context of spin (c.f., Eq.~(\ref{eq:kronecker})). 
We see that $\hat{M}_{r\rightarrow g}$ provides the desired operation of changing a red state to a green state
\begin{eqnarray}
\hat{M}_{r\rightarrow g}\left|r\right\rangle & \dot{=} & \left(\begin{array}{ccc} 0 & 0 & 0 \\ 1 & 0 & 0\\ 0 & 0 & 0\end{array}\right)\left(\begin{array}{c} 1\\ 0 \\ 0\end{array}\right) = \left(\begin{array}{c} 0 \\ 1 \\ 0\end{array}\right) \rightarrow \left| g\right\rangle.
\end{eqnarray}
It is straightforward to verify that $\hat{M}_{r\rightarrow g}$ can be written in terms of the SU(3) generators as $\hat{M}_{r\rightarrow g} = \hat{t}^{(1)} - i\hat{t}^{(2)}$, while $\hat{M}_{g\rightarrow r} = \hat{t}^{(1)} + i\hat{t}^{(2)}$. There are six such matrices to represent the six possible color changes with each given by a matrix with a single nonzero element which lies away from the diagonal. The space of {\it all} $3\times 3$ matrices requires nine basis matrices, with the other three given by
\begin{eqnarray}
\hat{M}_{r\rightarrow r} & \dot{=} & \left(\begin{array}{ccc} 1 & 0 & 0 \\ 0 & 0 & 0\\ 0 & 0 & 0\end{array}\right),\nonumber\\
\hat{M}_{g\rightarrow g} & \dot{=} & \left(\begin{array}{ccc} 0 & 0 & 0 \\ 0 & 1 & 0\\ 0 & 0 & 0\end{array}\right),\nonumber\\
\hat{M}_{b\rightarrow b} & \dot{=} & \left(\begin{array}{ccc} 0 & 0 & 0 \\ 0 & 0 & 0\\ 0 & 0 & 1\end{array}\right).\label{eq:mrr}
\end{eqnarray}
However, SU(3) is the group of unitary matrices with determinant equal to one, so generators of this group must be traceless.\cite{note3a} The matrices in Eq.~(\ref{eq:mrr}) are not traceless, but the following linear combinations are,
\begin{eqnarray}
\hat{M}_{r\rightarrow r} - \hat{M}_{g\rightarrow g} & \dot{=} & \left(\begin{array}{ccc} 1 & 0 & 0 \\ 0 & -1 & 0\\ 0 & 0 & 0\end{array}\right) = 2\hat{t}^{(3)},\nonumber\\
\hat{M}_{g\rightarrow g} - \hat{M}_{b\rightarrow b}& \dot{=} & \left(\begin{array}{ccc} 0 & 0 & 0 \\ 0 & 1 & 0\\ 0 & 0 & -1\end{array}\right) = \sqrt{3}\hat{t}^{(8)} - \hat{t}^{(3)},\nonumber\\
\hat{M}_{r\rightarrow r} - \hat{M}_{b\rightarrow b} & \dot{=} & \left(\begin{array}{ccc} 1 & 0 & 0 \\ 0 & 0 & 0\\ 0 & 0 & -1\end{array}\right) = \sqrt{3}\hat{t}^{(8)} + \hat{t}^{(3)}.
\end{eqnarray}
That is, only two independent generators are required to form all three of these traceless diagonal matrices. Restricting attention to traceless generators thus results in a total of eight independent matrices. Physically, these matrices correspond to the mediating particles of the strong interaction which are known as {\it gluons}. The underlying SU(3) structure in the strong interaction therefore implies the existence of eight ``independent'' types of gluons. At a conceptual level, one often says that a gluon carries a color and a single ``anti color.'' In our toy example of building a ``gluon'' from $\hat{M}_{r\rightarrow g} = \left|g\right\rangle \left\langle r\right|$, the object $\hat{M}_{r\rightarrow g}$ must carry one unit of ``green'' and one unit of ``anti red.'' Since ``green'' and ``anti red'' refer to states, the actual color charge of this object is ${\bf Q} = {\bf Q}_{\overline{\scriptsize r}} + {\bf Q}_{\mbox{\scriptsize g}} = -{\bf e}^{(3)}$. In terms of color content, one can identify the vectors ${\bf Q}_{\mbox{\scriptsize r,g,b}}$ with the red, green and blue color states unambiguously. The careful reader should be cautioned that the term ``unit of red charge'' is sometimes used (somewhat misleadingly) as shorthand for the charge vector ${\bf Q}_{\mbox{\scriptsize r}}$.

\section{Interactions}
\subsection{Two-body terms}
In this section we explore effective interactions between two sources of color charge. It should be noted that the strong nuclear interaction is highly modified by quantum processes. A na\"{i}ve exploration of the corresponding classical Lagrangian leads to the prediction of a Coulomb-like potential.\cite{Rubakov} The interactions between sources and the mediating fields result in a confining potential at large separations which makes is effectively impossible to separate two quarks in experiments.\cite{Halzen} In what follows, we consider a reduced model in which the sources interact directly through phenomenological interactions which are motivated below.

It is possible to write down direct coupling between the color sources and the external color fields, somewhat analogous to the Zeeman interaction between spin and magnetic field considered in Sec.~\ref{sec:onespin}. Such interactions are difficult, if not practically impossible, to probe experimentally due to the confinement of color charge within color-neutral hadrons and mesons. Consequently, we begin our discussion of interactions at the level of two-body systems with two sources of color charge. It is possible that future experiments with cold atoms using {\it artificial} gauge fields could make such external fields relevant to experiment.\cite{PhysicsToday} Looking to the Heisenberg Hamiltonian in Eq.~(\ref{eq:heisenberg}) for inspiration, we can postulate the existence of an effective two-body interaction between two color sources of the form
\begin{eqnarray}
\hat{H}_{\mbox{\scriptsize 2b}} & = & J \sum_{\alpha = 1}^{8}\hat{t}^{(\alpha)}\otimes\hat{t}^{(\alpha)},\label{eq:2bod}
\end{eqnarray}
for some positive constant $J$. The overall minus sign in front of Eq.~(\ref{eq:heisenberg}) was necessary to ensure parallel spins have a lower interaction energy than antiparallel spins. A possible motivation for the overall structure of Eq.~(\ref{eq:2bod}) follows from considering how this interaction energy behaves for several simple situations. First, consider two arbitrary sources in the state, $\left|\psi\right\rangle = \left|\psi_{1}\right\rangle \otimes\left|\psi_{2}\right\rangle$. The interaction energy is given by the expectation value of Eq.~(\ref{eq:2bod})
\begin{eqnarray}
E_{\mbox{\scriptsize int}}[\psi] & = & \left\langle \psi \right| J\sum_{\alpha=1}^{8}\hat{t}^{(\alpha)}\otimes\hat{t}^{(\alpha)}\left|\psi\right\rangle.
\end{eqnarray}
One property of the tensor product is that $\left[\hat{A}\otimes\hat{B}\right]\left[\left|\chi\right\rangle\otimes\left|\phi\right\rangle\right] = \left(\hat{A}\left|\chi\right\rangle\right)\otimes\left(\hat{B}\left|\phi\right\rangle\right)$, so the interaction can be written
\begin{eqnarray}
E_{\mbox{\scriptsize int}}[\psi_{1},\psi_{2}] & = & J{\bf Q}_{1}\cdot{\bf Q}_{2},\label{eq:eint}
\end{eqnarray}
where the single-source charge components are given by $Q_{1}^{(\alpha)} \equiv \left\langle \psi_{1} \right|\hat{t}^{(\alpha)}\left|\psi_{1}\right\rangle$, $Q_{2}^{(\alpha)} \equiv \left\langle \psi_{2} \right|\hat{t}^{(\alpha)}\left|\psi_{2}\right\rangle$. Equation~(\ref{eq:eint}) is formally a generalization of the Coulomb potential in which the notion of ``like'' and ``opposite'' charges has been generalized to an inner product of color charge vectors in gauge space. By treating only the color degree of freedom, we are neglecting information about the variation of the interaction energy with spatial separation of sources. Solutions to the two-body problem in {\it classical} chromodynamics\cite{Rubakov} lead to a Coulomb-like ($J\propto R^{-1}$) potential whose sign is determined by the sign of ${\bf Q}_{1}\cdot{\bf Q}_{2}$. Quantum effects related to the nonlinear field equations modify this potential strongly, resulting in a confining potential which grows {\it linearly} with separation, $J\propto R$ at large separation. A fairly robust interpolation for the static quark-antiquark potential is provided by the {\it Cornell potential},\cite{Cornell}
\begin{eqnarray}
J(R) & = & \frac{\alpha}{R} - \beta R.\label{eq:cornell}
\end{eqnarray}
Regardless of the precise functional form of $J(R)$, the relative orientation of charge vectors fixes whether the force between two sources is repulsive or attractive. For ${\bf Q}_{1}\cdot{\bf Q}_{2} >0$, this interaction energy is lowered by increasing the separation, so the resulting force is repulsive. Conversely, if ${\bf Q}_{1}\cdot{\bf Q}_{2}<0$, the interaction energy is lowered by decreasing the separation between sources, and the resulting force is attractive. The above reasoning applies to {\it any} color states $\left|\psi_{1}\right\rangle$ and $\left|\psi_{2}\right\rangle$, but let us specialize to the pure red, green and blue states in Eq.~(\ref{eq:colorbasis}) where the only nonzero color charge components exist in the two-dimensional $(Q^{(3)},Q^{(8)})$ plane. From Figure~\ref{fig:colorspace}, it is clear that the inner product of a quark's color charge vector with that of its antiquark will be maximally negative. As with electrons and positrons, Eq.~(\ref{eq:eint}) predicts that particles and their antiparticles should attract via the strong interaction. 

However, the the vectorial nature of color charge allows for a much richer theory of interactions than exists for the case of scalar charge. Without loss in generality, consider ${\bf Q}_{\mbox{\scriptsize r}}$. The angular displacement between ${\bf Q}_{\mbox{\scriptsize r}}$ and either ${\bf Q}_{\mbox{\scriptsize g}}$ or ${\bf Q}_{\mbox{\scriptsize b}}$ is 120$^{\circ}$, and since $\cos120^{\circ} <0$, there will be an attractive force between the red and blue sources and between the red and green sources. Identical reasoning can be used for the other quarks to show that there is mutual attraction between any two of these quarks. This type of calculation is based on a simplified, effective interaction. Consequently, the predictions should not be trusted for accurate results given the qualitative changes to the interactions when dynamical gluons are included in the theory. However, it is worth noting the qualitative features of quarks being mutually attractive and a three-particle bound state seeming plausible at this level are consistent with reality. Furthermore, these types of calculations are essentially equivalent to the tree-level calculations one can perform in quantum chromodynamics that do not take into account quantum fluctuations.\cite{GriffithsParticles}

The Hamiltonian in Eq.~(\ref{eq:2bod}) acts on states of {\it two} color sources. In a system of three sources, this pairwise interaction would contribute to each of the three possible pairings, so the appropriate generalization of Eq.~(\ref{eq:2bod}) for the case of three sources is
\begin{eqnarray}
\hat{H}_{\mbox{\scriptsize pairs}} & = & J \sum_{\alpha = 1}^{8}\left[\hat{t}_{1}^{(\alpha)}\cdot\hat{t}_{2}^{(\alpha)} + \hat{t}_{1}^{(\alpha)}\cdot\hat{t}_{3}^{(\alpha)} + \hat{t}_{2}^{(\alpha)}\cdot\hat{t}_{3}^{(\alpha)}\right],\label{eq:2body}
\end{eqnarray}
where the appropriate identity operators not written explicitly. For example,
\begin{eqnarray}
 \hat{t}_{1}^{(\alpha)}\cdot\hat{t}_{3}^{(\alpha)} & \equiv & \hat{t}^{(\alpha)}\otimes I\otimes \hat{t}^{(\alpha)}.
 \end{eqnarray}
We consider a simple product state corresponding to all three basis color states,
\begin{eqnarray}
\left|\psi\right\rangle & = & \left|r\right\rangle \otimes  \left|g\right\rangle \otimes  \left|b\right\rangle,\label{eq:psi00}
\end{eqnarray}
which has no net color charge. The state in Eq.~(\ref{eq:psi00}) state mimics that of a baryon, which is also colorless, since the net color charge is zero. It should be noted that actual baryons are in a color {\it singlet} configuration, which is a totally antisymmetric superposition state analogous the the spin singlet state $\left|0,0\right\rangle = \frac{1}{\sqrt{2}}\left(\left|\uparrow\right\rangle \left|\downarrow\right\rangle  - \left|\downarrow\right\rangle \left|\uparrow\right\rangle \right)$. Explicitly,
\begin{eqnarray}
\left|\psi_{\mbox{\scriptsize singlet}}\right\rangle & = & \frac{1}{\sqrt{6}}\left(\left|rgb\right\rangle -\left|rbg\right\rangle +\left|gbr \right\rangle \right.\nonumber\\
& & -\left.\left|grb\right\rangle + \left|brg\right\rangle - \left|bgr\right\rangle\right),\label{eq:singlet}
\end{eqnarray}
where we have employed shorthand notation $\left|rgb\right\rangle \equiv \left|r\right\rangle \otimes\left| g\right\rangle\otimes\left|b\right\rangle$, and similarly for other terms. The significance of naturally-occurring bound states employing the singlet color configuration cannot be understated, and this is intimately related to net color charge being essentially unobservable in realistic experiments. Indeed the singlet configuration is invariant with respect to gauge transformations in $\mbox{SU}(3)$, whereas the simple product state $\left|\psi\right\rangle$ is not.\cite{GriffithsParticles} By using superposition to take full advantage of the underlying symmetry, we will see that the singlet configuration actually possesses a lower ground state energy than Eq.~(\ref{eq:psi00}) for our choice of effective interactions. Because gauge transformations amount to a redefinition of color charge components,\cite{Boozer} gauge-invariant dynamics require net color charge itself to be effectively unobservable, somewhat like the scalar and vector potentials in electromagnetism. Unless otherwise noted, we will consider the product state $\left|\psi\right\rangle$ as the initial state for simplicity. This theoretical choice is for pedagogical simplicity, but the resulting calculations are not entirely decoupled from experimental reality. While actual quarks are found in the singlet state, nothing forbids simulating non-singlet states using cold atoms and artificial gauge fields.~\cite{PhysicsToday} The corresponding interaction energy takes the form
\begin{eqnarray}
E_{\mbox{\scriptsize int}}[\psi_{1},\psi_{2},\psi_{3}] & = & J\left({\bf Q}_{1}\cdot{\bf Q}_{2} + {\bf Q}_{1}\cdot{\bf Q}_{3}+{\bf Q}_{2}\cdot{\bf Q}_{3}\right).\label{eq:pairint}
\end{eqnarray}
Let us calculate explicitly the interaction energy for the state $\left|\psi\right\rangle = \left|r\right\rangle \otimes \left|g\right\rangle\otimes \left|b\right\rangle$. Eqs.~(\ref{eq:pairint}) and~(\ref{eq:rgbcharge1})--(\ref{eq:rgbcharge3}) give
 \begin{eqnarray}
 E_{\mbox{\scriptsize int}}[r,g,b] & = & J\left({\bf Q}_{\mbox{\scriptsize r}}\cdot{\bf Q}_{\mbox{\scriptsize g}} + {\bf Q}_{\mbox{\scriptsize r}}\cdot{\bf Q}_{\mbox{\scriptsize b}} + {\bf Q}_{\mbox{\scriptsize g}}\cdot{\bf Q}_{\mbox{\scriptsize b}}\right),\nonumber\\
 & = & -\frac{J}{2}.\label{eq:prode}
 \end{eqnarray}
The minus sign indicates in Eq.~(\ref{eq:prode}) indicates an attractive potential. Now consider the singlet state in Eq.~(\ref{eq:singlet}). One can show that the singlet does correspond to a lower energy than the simple product state for interactions given by Eq.~(\ref{eq:2body}). For a superposition state such as $\left|\psi_{\mbox{\scriptsize singlet}}\right\rangle$, we must use Eq.~(\ref{eq:2body}) rather than the ``classical'' expression given by Eq.~(\ref{eq:pairint}). To see this, let us explore a single term that arises as $\hat{H}_{\mbox{\scriptsize pairs}}$ acts on the first term of the singlet state,
 \begin{eqnarray}
\hat{t}^{(1)}\otimes\hat{t}^{(1)}\otimes \hat{I}\left( \left| r\right\rangle\otimes\left|g\right\rangle\otimes \left|b\right\rangle\right) & = & \frac{1}{4}\left|g\right\rangle\otimes\left|r\right\rangle\otimes\left|b\right\rangle.\label{eq:singterm}
\end{eqnarray}
If we were considering the product state $\left|rgb\right\rangle$, such a term would vanish upon taking the inner product with $\left\langle rgb\right|$. However, the singlet state also contains a term $\frac{1}{\sqrt{6}}\left|grb\right\rangle$, so Eq.~(\ref{eq:singterm}) makes a nonzero contribution to the energy. The Gell-Mann matrices are fairly sparse, so most such cross terms still vanish. However, there are a number of terms to keep track of, and after either a fairly careful accounting or a few minutes with a computer algebra package, one finds
\begin{eqnarray}
E_{\mbox{\scriptsize int}}[\psi_{\mbox{\scriptsize singlet}}] & = & -2J,
\end{eqnarray}
which is, as advertised, a lower energy than obtained for the simple product state in Eq.~(\ref{eq:psi00}). One can think of a state such as $\left|rgb\right\rangle$ being somewhat ``classical'' in the sense that a measurement of $Q^{(3)}$ or $Q^{(8)}$ on any of the three sources would always return the same answer, since each quark is in a simultaneous eigenstate of both $\hat{Q}^{(3)}$ and $\hat{Q}^{(8)}$. The singlet state gives an expectation value of zero for each charge component for each source. However, as with any quantum mechanical observable, an individual measurement of $\hat{Q}^{(3)}$ and $\hat{Q}^{(8)}$ would return one of the operator's eigenvalues as a result. The measurement forces one source to ``pick'' one of the three positive color charge eigenstates. Because the singlet state is entangled, the measurement of source one also affects the state of sources two and three. Entanglement is sometimes viewed as one of the fundamentally ``quantum'' features systems can exhibit, and the singlet state is in some sense more inherently ``quantum'' than the product state. The singlet is also possesses a higher degree of symmetry than the product state, being antisymmetric with respect to the exchange of sources, $\left|\psi_{\mbox{\scriptsize singlet}}(r,b,g)\right\rangle =-\left|\psi_{\mbox{\scriptsize singlet}}(r,g,b)\right\rangle$. One may also show that the singlet is invariant with respect to $\mbox{SU}(3)$ gauge transformations which redefine the color charge components.~\cite{Boozer}

A similar situation arises with the ground state of the one-dimensional Heisenberg antiferromagnet, which is composed of $N$ spin-$\frac{1}{2}$ degrees of freedom with interactions between nearest neighbors described by the Hamiltonian
\begin{eqnarray}
\hat{H}_{\mbox{\scriptsize Heisenberg}} & = & J\sum_{j=1}^{N-1}\left[\hat{S}_{j}^{x}\hat{S}_{j+1}^{x} + \hat{S}_{j}^{y}\hat{S}_{j+1}^{y} + \hat{S}_{j}^{z}\hat{S}_{j+1}^{z}\right],
\end{eqnarray}
where $J>0$ ensures antiferromagnetic interactions. A plausibly simple candidate for a ground state is the N\'{e}el state, $\left|\uparrow\downarrow\uparrow\cdots\downarrow\uparrow\downarrow\right\rangle$. One reason this cannot be the ground state is found by observing the inversion symmetry of $\hat{H}_{\mbox{\scriptsize Heisenberg}}$ is not respected by this state. That is, another equally good--but equally problematic--candidate is $\left|\downarrow\uparrow\downarrow\cdots\uparrow\downarrow\uparrow\right\rangle$. A more glaring shortcoming of the N\'{e}el state is that while it minimizes the energetic contribution from the $\hat{S}_{j}^{z}\hat{S}_{j+1}^{z}$ terms, it is not even an eigenstate of the full Hamiltonian. To see this, note that the first two terms in Eq.~(\ref{eq:heisenberg}) may be recast as
\begin{eqnarray}
\hat{S}_{j}^{x}\hat{S}_{j+1}^{x} + \hat{S}_{j}^{y}\hat{S}_{j+1}^{y} & = &  \hat{S}_{j}^{+}\hat{S}_{j+1}^{-} + \hat{S}_{j}^{-}\hat{S}_{j+1}^{+},
\end{eqnarray}
where $\hat{S}_{j}^{\pm} \equiv \hat{S}_{j}^{x} \pm i\hat{S}^{y}_{j}$ are the spin raising/lowering operators.~\cite{GriffithsQM} Each term raises (lowers) a particular spin and lowers (raises) its right neighbor. This process is analogous to the example for the three-source $\mbox{SU}(3)$ color charge system considered above in the singlet state. However, the equivalent ``singlet'' state for the case of $N$ spins is substantially more complicated than that of three $\mbox{SU}(3)$ color charges, and a fairly elaborate technique known as the Bethe ansatz\cite{Giamarchi} must be used to obtain the ground state.
\subsection{Three-body interactions}
Hadrons, such as protons and neutrons, are bound states of three quarks. In addition to the two-body interactions, a system of three particles could experience legitimate, three-body interactions which cannot be decomposed into a sum of pairwise interactions between pairs of particles. Indeed, the full theory of quantum chromodynamics leads to {\it many-body} interactions between sources of color charge. The three-body interactions considered here amount to an effective interaction arising from the dynamics of mediating gluons which we neglect in this simplified description. Including dynamical quantum mechanical degrees of freedom for these mediating fields amounts to a legitimate quantum {\it field} theory and is necessarily much more complex than the idealized model presented here. In quantum chromodynamics, three-body interaction terms between the mediating gluon fields appear in the defining Lagrangian of the theory,~\cite{Peskin} which should result in effective three-body interactions between sources at the level of the effective interactions we consider. 

In a system of three sources, the general structure of three-body interactions is captured by the following Hamiltonian,
\begin{eqnarray}
\hat{H}_{\mbox{\scriptsize 3b}} & = & \sum_{\alpha,\beta,\gamma} g_{\alpha\beta\gamma}\hat{t}^{(\alpha)}\otimes\hat{t}^{(\beta)}\otimes\hat{t}^{(\gamma)},\label{eq:h3b}
\end{eqnarray}
where $g_{\alpha\beta\gamma}$ is some tensor of coefficients. The interactions in quantum chromodynamics, or any gauge theory, are highly constrained by the symmetries of the underlying gauge group.~\cite{Zee} At the level of constructing an effective interaction Hamiltonian, this requirement of gauge invariance means that the Hamiltonian must commute with each {\it total} component of color charge $t^{(\alpha)}_{\mbox{\scriptsize total}}$, where
\begin{eqnarray}
t^{(\alpha)}_{\mbox{\scriptsize total}} & \equiv & \hat{t}_{1}^{(\alpha)} + \hat{t}_{2}^{(\alpha)} + \hat{t}_{3}^{(\alpha)},\nonumber\\
& = & \hat{t}^{(\alpha)}\otimes \hat{I}\otimes\hat{I} + \hat{I}\otimes\hat{t}^{(\alpha)}\otimes \hat{I} + \hat{I}\otimes\hat{I}\otimes\hat{t}^{(\alpha)}.
\end{eqnarray}
While searching blindly for suitable operators might appear to be a daunting task, the structure of the underlying gauge group provides two appealing possibilities for the form of $g_{\alpha\beta\gamma}$. The ${\mbox{SU}(3)}$ generators in Eq.~(\ref{eq:gellmann}) satisfy a set of relations which serve to define the so-called {\it structure constants} $d^{\alpha\beta\gamma}$ and $f^{\alpha\beta\gamma}$,
\begin{eqnarray}
\left[\hat{t}^{(\alpha)}, \hat{t}^{(\beta)}\right] & = & i\sum_{\gamma}f^{\alpha \beta \gamma}\hat{t}^{(\gamma)},\\
\left\{ \hat{t}^{(\alpha)},\hat{t}^{(\beta)}\right\} & = & \frac{1}{3}\delta^{\alpha \beta} + \sum_{\gamma}d^{\alpha\beta\gamma}\hat{t}^{(\gamma)}.
\end{eqnarray}
where $\left[A,B\right] \equiv AB-BA$ is the commutator and $\left\{A,B\right\} = AB+BA$ is the anticommutator. The Kronecker delta is defined by $\delta^{ab} = 1$ for $a=b$ and $\delta^{ab} = 0$ otherwise. By virtue of the antisymmetry (symmetry) of the commutator (anticommutator) with respect to indices, one may verify that the $d^{\alpha\beta\gamma}$ are totally symmetric while the $f^{\alpha\beta\gamma}$ are totally antisymmetric with respect to the indices $\alpha$, $\beta$, $\gamma$. There are a total of $8^{3} = 512$ possible index combinations for each, but most turn out to be zero. The following values are obtained\cite{ZeeGroup}
\begin{eqnarray}
d^{118} = d^{228} = d^{338} = -d^{888} & = & \frac{1}{\sqrt{3}},\\
d^{146} = d^{157} = d^{256} = d^{344} = d^{355} & = & \frac{1}{2},\\
d^{247} = d^{366} = d^{377} & = & -\frac{1}{2},\\ 
d^{448} = d^{558} = d^{668} = d^{778} & = & -\frac{1}{2\sqrt{3}},
\end{eqnarray}
\begin{eqnarray}
f^{123} & = & 1,\\
f^{147} = f^{246} =  f^{257}  = f^{345} & = & \frac{1}{2},\\
f^{156} = f^{367} & = & -\frac{1}{2},\\
f^{458} = f^{678} & = & \frac{\sqrt{3}}{2}.
\end{eqnarray}
Aside from entries which may be obtained from symmetry via cyclic permutations, $d^{\alpha\gamma\beta} = d^{\beta\alpha\gamma} = d^{\alpha\beta\gamma}$, $f^{\gamma\beta\alpha} = f^{\alpha\gamma\beta} = f^{\beta\alpha\gamma} = -f^{\alpha\beta\gamma}$, and $f^{\gamma\alpha\beta} = f^{\beta\gamma\alpha} = f^{\alpha\beta\gamma}$, all other components are zero.

One may compute gauge scalars (or pseudoscalars) from these structure constants of the forms 
\begin{eqnarray}
\sum_{\alpha,\beta,\gamma}d^{\alpha\beta\gamma}\hat{t}^{(\alpha)}\hat{t}^{(\beta)}\hat{t}^{(\gamma)}, \;\;\;\;\;\; \sum_{\alpha,\beta,\gamma}f^{\alpha\beta\gamma}\hat{t}^{(\alpha)}\hat{t}^{(\beta)}\hat{t}^{(\gamma)}.\label{eq:inv}
\end{eqnarray}
 The symmetric scalar resembles the component form of the ordinary scalar products between two three-dimensional vectors,
\begin{eqnarray}
{\bf A}\cdot{\bf B} & = & \sum_{i,j}\delta_{ij}A_{i}B_{j},\\
\end{eqnarray}
which is invariant with respect to three-dimensional rotations. In the same sense, the quantities in Eq.~(\ref{eq:inv}) are invariant with respect to $\mbox{SU(3)}$ gauge transformations. Interaction terms must be invariant with respect to $\mbox{SU(3)}$ gauge transformations, commuting with each of the $\hat{t}^{(\alpha)}$. Accordingly the expressions in~(\ref{eq:inv}) provide suitable candidates for interaction terms involving three $\hat{t}^{(\alpha)}$ operators which are inspired by the gauge group invariants. However, the quantities in Eq.~(\ref{eq:inv}) are simple matrix products. It can be shown that each is proportional to the $3\times 3$ identity matrix. When promoted to three-body interaction terms, this triple matrix product is replaced by a triple Kronecker product, creating a $3^{3}\times 3^{3}$ matrix which is {\it not} proportional to the identity matrix but which {\it does} commute with each of the $\hat{t}^{(\alpha)}$. This line of reasoning also motivates the two-body interaction proposed in Eq.~(\ref{eq:2body}). By squaring and summing the components of color charge, we arrive at an expression analogous to the squared spin operator in Eq.~(\ref{eq:s2single}),
\begin{eqnarray}
\sum_{\alpha = 1}^{8}\left[\hat{t}^{(\alpha)}\right]^{2} & = & \frac{4}{3}\hat{I}.\label{eq:casimir2}
\end{eqnarray}
By replacing the matrix product in Eq.~(\ref{eq:casimir2}) by a tensor product, we obtain the two-body interaction in Eq.~(\ref{eq:2body}) which commutes with each $\hat{t}^{(\alpha)}_{\mbox{\scriptsize total}}$ for the case of two sources. Incidentally, this structure immediately implies the conservation of each total charge component individually, though the charge components for each source can vary individually. Lastly, it should be noted that the combination involving the asymmetric structure constants should also contain factors which depend on the spin on the sources.~\cite{PLB2001} Indeed, the gauge scalar obtained from $f^{\alpha\beta\gamma}$ is somewhat similar to a pseudoscalar obtained from an ordinary vector cross product. Our focus is on exploring dynamics with minimum complications, so we will adopt both terms as possible forms for three-body interactions. 
\section{Dynamics}
\label{sec:dynamics}
Our goal in this section is to repeat the basic steps from Sec.~\ref{sec:spin} for a three-quark product state having the form
 \begin{eqnarray}
 \left|q_{1},q_{2},q_{3}\right\rangle & = & \left|q_{1}\right\rangle \otimes \left|q_{2}\right\rangle \otimes \left|q_{3}\right\rangle,\label{eq:initstate}
 \end{eqnarray}
 with time evolution generated by the Hamiltonian
 \begin{eqnarray}
 \hat{H} & = & \hat{H}_{\mbox{\scriptsize pairs}} + \hat{H}_{\mbox{\scriptsize 3-body}},\\
 & = & J\sum_{\alpha}\left[\hat{t}^{(\alpha)}_{1}\cdot\hat{t}^{(\alpha)}_{2} + \hat{t}^{(\alpha)}_{1}\cdot\hat{t}^{(\alpha)}_{3} + \hat{t}^{(\alpha)}_{2}\cdot\hat{t}^{(\alpha)}_{3}\right]\nonumber\\
&+ & V\sum_{\alpha,\beta,\gamma}g_{\alpha\beta\gamma}\hat{t}^{(\alpha)}\otimes\hat{t}^{(\beta)}\otimes \hat{t}^{(\gamma)},\label{eq:fullham}
 \end{eqnarray}
 where $\Delta \equiv \frac{V}{J}$ is a dimensionless parameter to control the relative strength of the three-body interaction terms. The task at hand is to solve the time-dependent Schr\"{o}dinger equation for a given initial state and compute the expectation values of observables (i.e., color charge components). Before proceeding to investigate the dynamics, we must obtain explicit representations of the initial state in Eq.~(\ref{eq:initstate}) and relevant operators in Eq.~(\ref{eq:fullham}) which are suitable for numerical or analytic analysis.
\subsection{Two-body interactions ($V = 0$)}
Let us first consider the case $V = 0$ so that only two-body interactions are present. This case admits an analytic solution for the color charge components through straightforward, if tedious steps. Alternatively, the numerical approach developed for three-body interactions in the next section is easily adapted to the case of two body interactions. The general state of Eq.~(\ref{eq:initstate}) is a tensor product of three color states and is represented by a $3^{3} = 27$-dimensional column vector
 \begin{eqnarray}
 \left|\psi_{0}\right\rangle & \dot{=} & \left(\begin{array}{c} c_{1}\\c_{2}\\ \vdots\\c_{27}\end{array}\right),\label{eq:initco}
 \end{eqnarray}
 where the $c_{i}$ are complex numbers. For a simple product state of the three basis vectors, $\left|\psi_{0}\right\rangle = \left| r\right\rangle \otimes \left| g\right\rangle \otimes \left|b\right\rangle$, one finds $c_{6} = 1$, with $c_{i} = 0$ for $i\neq 6$. In principle, it is possible to work out the action of the two-body Hamiltonian operator in Eq.~(\ref{eq:2body}) on the general state $\left|\chi\right\rangle$, obtaining a set of coupled differential equations for the coefficients $c_{i}$ via the Schr\"{o}dinger equation
 \begin{eqnarray}
 i\hbar \frac{\partial}{\partial t}\left|\psi\right\rangle & = & \hat{H}_{\mbox{\scriptsize pairs}}\left|\psi\right\rangle.\label{eq:schrodinger}
 \end{eqnarray}
Upon substituting Eq.~(\ref{eq:initco}) into Eq.~(\ref{eq:schrodinger}), one may use the explicit forms of the Gell-Mann matrices in Eq.~(\ref{eq:gellmann}) to obtain the action of $\hat{H}_{\mbox{\scriptsize pairs}}$ on $\left|\psi\right\rangle$. The resulting 27 equations split into decoupled sets, so that only those equations involving $c_{6}$ are nontrivial. All others involve only quantities which have been initialized to zero and will never evolve into nonzero values. In what follows, we work in units where $\hbar\rightarrow 1$ for brevity. This linear system can be solved explicitly for the nontrivial coefficients using the initial conditions $c_{6}(0) = 1$ with all others zero, giving
\begin{eqnarray}
c_{6}(t) & = & \frac{1}{6}e^{-iJt}- \frac{1}{6}e^{2iJt},\label{eq:c6t}\\
c_{8}(t) & = & \frac{1}{6}e^{-iJt}+\frac{1}{6}e^{2iJt} - \frac{1}{3}e^{\frac{i}{2}Jt},\label{eq:c8t}\\
c_{12}(t) & = & \frac{1}{6}e^{-iJt}+ \frac{1}{6}e^{2iJt} - \frac{1}{3}e^{\frac{i}{2}Jt},\label{eq:c12t}\\
c_{16}(t) & = & \frac{1}{6}e^{-iJt}-\frac{1}{6}e^{2iJt},\label{eq:c16t}\\
c_{20}(t) & = & \frac{1}{6}e^{-iJt} - \frac{1}{6}e^{2iJt},\label{eq:c20t}\\
c_{22}(t) & = & \frac{1}{6}e^{iJt} + \frac{1}{6}e^{2iJt} + \frac{2}{3}e^{\frac{i}{2}Jt}.\label{eq:c22t}
\end{eqnarray}
Expectation values of charge components follow from generalizing Eq.~(\ref{eq:chargecomp}) to the case of three sources. For example, the charge components of the first source are given by
\begin{eqnarray}
Q_{1}^{(\alpha)}(t) & = & \left\langle \psi (t)\right| \hat{t}^{(\alpha)}\otimes \hat{I}\otimes \hat{I}\left|\psi(t)\right\rangle,
\end{eqnarray}
 and depend on the nonzero $c_{i}(t)$. Here ${\bf Q}_{1,2,3}(t)$ is the color charge vector of the source which was initially red, green, and blue, respectively. Using the explicit solutions for $c_{i}(t)$. 
these expectation values reduce to
\begin{eqnarray}
Q_{1}^{(3)}(t) = -Q_{2}^{(3)}(t) & = & \frac{1}{2}f(t),\label{eq:2bodsol1}\\
Q_{3}^{(3)}(t) & = & 0,\\
Q_{1}^{(8)}(t) = Q_{2}^{(8)}(t) & = & \frac{1}{2\sqrt{3}}f(t),\\
Q_{3}^{(8)}(t) & = & -\frac{1}{\sqrt{3}}f(t),\label{eq:2bodsol2}
\end{eqnarray}
where $f(t) \equiv \frac{1}{3}\left(1 + 2\cos\left[\frac{3}{2}Jt\right]\right)$. Equations~(\ref{eq:2bodsol1})--(\ref{eq:2bodsol2}) predict oscillations in the charge components, which are most easily visualized by an animation of the evolution in $(Q^{(3)},Q^{(8)})$ space as discussed in Sec.~\ref{sec:visualization}. Already one finds a qualitatively different type of behavior than occurs in electrostatics in which the charge is a scalar which does not possess any dynamics.~\cite{note6} These oscillations describe a sort of flip-flop behavior with the color sliding along the directions defined by Figure~\ref{fig:colorspace} and switching between ``positive'' and ``negative,'' where ``positive'' corresponds to parallel to the color's initial direction and ``negative'' being antiparallel. Such nontrivial charge dynamics also persists in the corresponding classical field theory, as has been demonstrated explicitly~\cite{Boozer} for the case of $\mbox{SU}(2)$ dynamics in which the sources move under the influence of interaction forces between the individual sources. From Eqs.~(\ref{eq:2bodsol1})--(\ref{eq:2bodsol2}) one sees that while individual components oscillate, the total charge components always add to zero for the three-source system.\\[2ex]

\subsection{Three-body interactions ($V \neq 0$)}

The case in which three-body interactions are included is significantly more complex, so we employ a numerical approach to make the treatment as accessible as possible. This numerical approach also provides an alternative method to investigate the two-body interactions considered in the previous section. A \textsc{Jupyter} notebook which makes use of several convenient \textsc{Numpy} functions is included in the supplementalary material~\cite{supp} and allows the reader to recreate all cases studied here by simply changing parameters. To reduce the number of free parameters, we restrict attention to the antisymmetric three-body interactions, $g_{\alpha\beta\gamma}\rightarrow f^{\alpha\beta\gamma}$, in what follows.

The basic steps required to obtain $Q_{1,2,3}^{(\alpha)}(t)$ are to (1) build a matrix representation of the Hamiltonian generating time evolution, (2) construct an explicit vector representation of the initial state $\left|\psi_{0}\right\rangle$, (3) solve the Schr\"{o}dinger equation to obtain $\left|\psi(t)\right\rangle$, and (4) compute appropriate inner products of the form $\left\langle \psi(t)\right|\hat{O}\left|\psi(t)\right\rangle$ where $\hat{O}$ is some 3-body operator corresponding to a component of one source's color charge.  Particularly helpful with steps (1) and (4) is the \textsc{Numpy} function \texttt{kron(A,B)} which computes the Kronecker product of two matrices, \texttt{A} and \texttt{B}. The Kronecker product provides an explicit representation of the abstract tensor product in Eqs.~(\ref{eq:initstate}) and (\ref{eq:fullham}), so much of the tedious work can be performed behind the scenes, leading to a fairly compact program. The interested reader can find an explicit scheme for constructing appropriate single-site matrices in multi-site systems in Ref.~\onlinecite{CandelaAJP} which also generalizes to operators which cannot be decomposed into a Kronecker product. 

For an $N$-dimensional Hamiltonian $\hat{H}$ with energy eigenvalues $\epsilon_{n}$ and corresponding eigenstates $\left|\phi_{n}\right\rangle$, hermiticity guarantees that any state $\left|\psi\right\rangle$ may be written as a linear combination of the eigenstates
\begin{eqnarray}
\left|\psi_{0}\right\rangle & = & \sum_{n=1}^{N}c_{n}\left|\phi_{n}\right\rangle,
\end{eqnarray}
where the coefficients $c_{n} = \left\langle \phi_{n}\right|\left. \psi_{0}\right\rangle$ represent the overlap between the state $\left|\psi_{0}\right\rangle$ and the $n^{\mbox{\scriptsize th}}$ eigenstate of $\hat{H}$. Since the eigenstates have trivial time evolution $\left|\phi_{n}(t)\right\rangle = e^{-i\epsilon_{n}t}\left|\phi_{n}\right\rangle$ as stationary states, the full time-dependence of an arbitrary state $\left|\psi_{0}\right\rangle$ can be written as
\begin{eqnarray}
\left|\psi(t)\right\rangle & = & \sum_{n}c_{n}e^{-i\epsilon_{n}t}\left|\phi_{n}\right\rangle.
\end{eqnarray}
The built-in \textsc{Numpy} routine \texttt{w,v = eigh(H)} provides a list of eigenvalues \texttt{w} and matrix \texttt{v} whose columns are the corresponding eigenvectors of a (Hermitian) matrix \texttt{H}. Accordingly, we use this diagonalization procedure to solve the Schr\"{o}dinger equation numerically.

We employ three-dimensional arrays \texttt{A[i,j,k]} to store the coefficients or $f^{\alpha\beta\gamma}$ and $d^{\alpha\beta\gamma}$ as well as the Gell-Mann matrices. For the latter, we define an array \texttt{ts} such that \texttt{ts[:,:,n]} is the $n^{\mbox{\scriptsize th}}$ Gell-Mann matrix.~\cite{note7} By indexing these matrices, the sums in Eq.~(\ref{eq:fullham}) can be easily written as an unrestricted sum and coded as a series of nested loops.

Operators corresponding to one-body observables are constructed by applying nested instances of \texttt{kron()} to a Gell-Mann matrix and two factors of the identity matrix (stored as \texttt{t0} in the program),
\begin{eqnarray}
\hat{t}^{(2)}\otimes \hat{I}\otimes\hat{I} \rightarrow \mbox{\texttt{kron(ts[:,:,1],kron(t0,t0))}}.
\end{eqnarray}
\begin{figure}
\begin{center}
\includegraphics[totalheight=19.5cm]{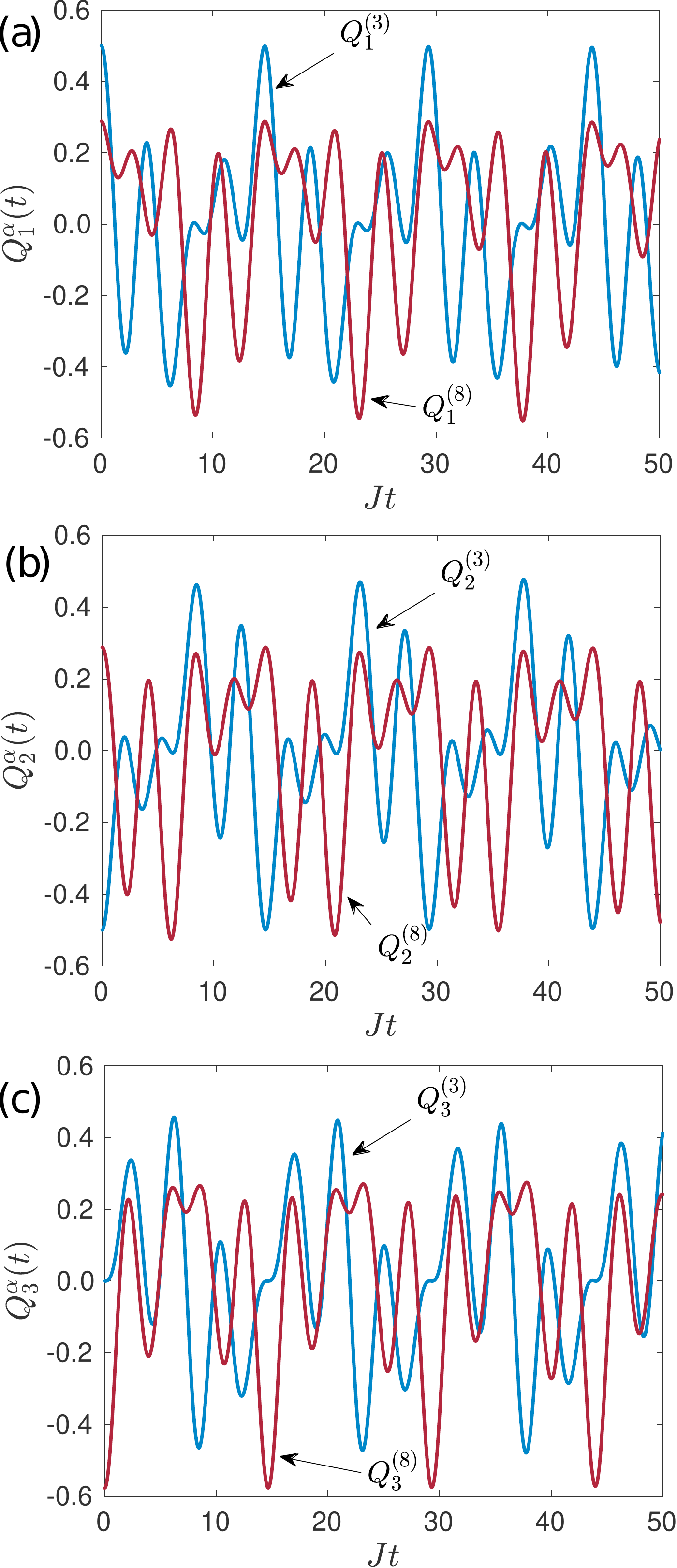}
\caption{Color charge components as a function of time for $\Delta = 0.5$ for (a) $Q_{1}^{(\alpha)}(t)$; (b) $Q_{2}^{(\alpha)}(t)$; (c) $Q_{3}^{(\alpha)}(t)$; only the $\alpha = 3,8$ components become nonzero, so other components are not shown.}
\label{fig:colortime}
\end{center}
\end{figure}
Figure~\ref{fig:colortime} depicts typical charge component dynamics for $V \neq 0$. We still observe periodic variation in charge components, though it is somewhat more complicated than the dynamics observed for $V = 0$ (c.f, Eqs.~(\ref{eq:2bodsol1})--(\ref{eq:2bodsol2})). In addition to the sinusoidal modulation of individual charge components, the charge components also rotate in the $(Q^{(3)},Q^{(8)})$ plane. Since the interactions contained in Eq.~(\ref{eq:fullham}) commute with the total color charge operators, $\hat{t}^{(\alpha)}_{\mbox{\scriptsize total}}$, we do not show the other components, as they are fixed to zero. As with two-body interactions, all total charge components for the three-source system sum to zero at all times, and the three angles separating the charge vectors are fixed at 120$^{\circ}$. Time series plots such as those in Fig.~\ref{fig:colortime} are not the most transparent depictions of the dynamics, and a clearer visualization scheme is presented in Sec.~\ref{sec:visualization}. It worth emphasizing that the equivalent situation in electromagnetism is two charged particles fixed in place where no dynamics arise.   

The patient reader may verify that applying the same tedious steps as for the two-body interactions also leads to a closed-form, analytic solution for the charge components with
\begin{eqnarray}
Q_{3}^{(3)}(t) & = & \frac{2\sin\left(\frac{\sqrt{3}Vt}{4}\right)}{3\sqrt{3}}\left[ \cos\left(\frac{\sqrt{3}Vt}{4}\right)-\cos\left(\frac{3Jt}{2}\right)\right],\label{eq:3bodsol1}\\
Q_{3}^{(8)}(t) & = & -\frac{2\cos\left(\frac{\sqrt{3}Vt}{4}\right)}{3\sqrt{3}}\left[\cos\left(\frac{\sqrt{3}Vt}{4}\right)+\cos\left(\frac{3Jt}{2}\right)\right]\nonumber\\
& & \left.+\frac{1}{3\sqrt{3}}\right].\label{eq:3bodsol2}
\end{eqnarray}
Other source components can be obtained by applying successive rotations of 120$^{\circ}$ to the vector with components $(Q_{3}^{(3)}(t),Q_{3}^{(8)}(t))$. Computer algebra software such as \textsc{Mathematica} is quite useful in dealing with the significant algebra involved in obtaining Eqs.~(\ref{eq:3bodsol1})--(\ref{eq:3bodsol2}). 
\begin{figure}
\begin{center}
\includegraphics[totalheight=7cm]{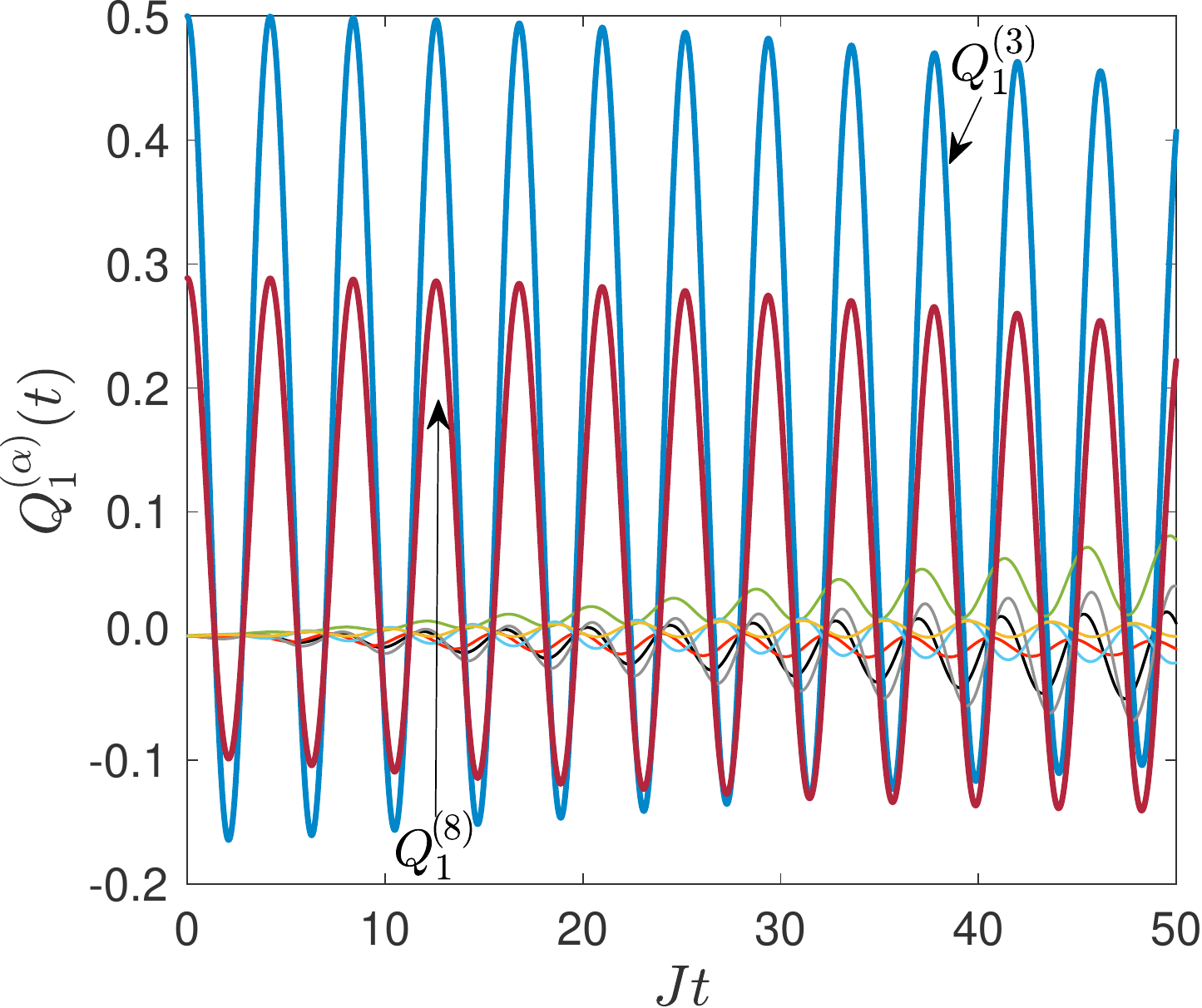}
\caption{Using random couplings in the three-body term $\hat{H}_{\mbox{\scriptsize 3-body}}$ with $\xi \in [0,0.01]$ allows the charge components with $\alpha \neq 3,8$ to gradually become nonzero. The time required for these other components to become comparable in size to $Q^{(3,8})$ decreases with increasing $\xi$.}
\label{fig:random3bod}
\end{center}
\end{figure}
One may also explore whether there are any observable consequences for not using the particular choices for three-body interaction coefficients of $f^{\alpha\beta\gamma}$ and $d^{\alpha\beta\gamma}$. With the numerical approach, one is not bound to study highly symmetric and analytically tractable situations. Figure~\ref{fig:random3bod} depicts the numerical solution for the case in which the coefficients of the three-body interaction in Eq.~(\ref{eq:fullham}) are replaced by random numbers $f^{\alpha\beta\gamma}\rightarrow \xi$. In this case we draw from a uniform distribution $\xi \in [0,0.01]$. For short times, the system essentially follows the two-body solution. Gradually, however, the random three-body couplings cause the other charge components to become nonzero. The reader may verify using the provided program that using only two-body couplings with random values is {\it not} sufficient to ``turn on'' the components with $\alpha \neq 3,8$.

\subsection{Visualization}
\label{sec:visualization}

\begin{figure}
\begin{center}
\includegraphics[totalheight=8.65cm]{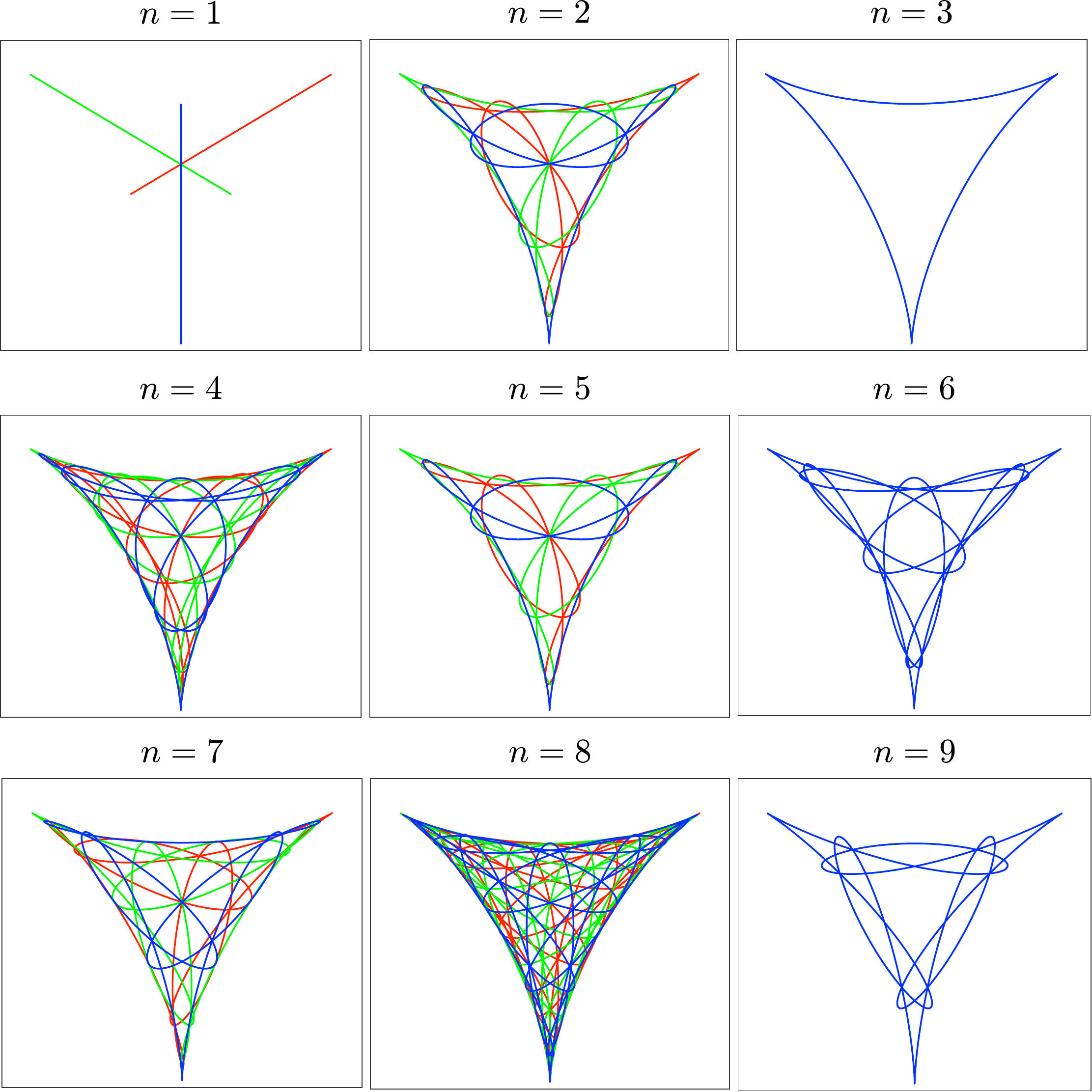}
\caption{Periodic orbits in the $(Q^{(3)},Q^{(8)})$ plane for various values of $n$ and $\Delta_{n} = 2\sqrt{3}/n$. The case where $n$ is a multiple of 3 is particularly degenerate, with all three source vectors moving along the same closed curve. The initial state of each source (red, blue, green) fixes the color of each trajectory.}
\label{fig:periodic}
\end{center}
\end{figure}

\begin{widetext}

\begin{figure}
\begin{center}
\includegraphics[totalheight=3.50cm]{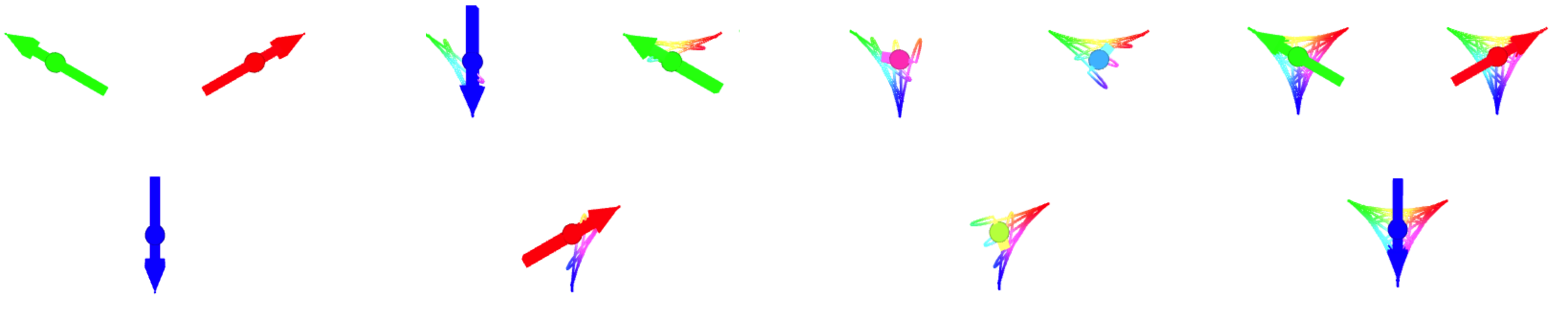}
\caption{Snapshots of the \textsc{VPython} animation for periodic orbit. The color charges evolve under a periodic orbit with $n=19$. Each arrow direction follows the orientation of that source's color charge vector in $(Q^{(3)},Q^{(8)})$ space. The instantaneous color of each source can be inferred by comparing the direction of the corresponding charge vector to the orientations in Fig.~\ref{fig:colorspace}.}
\label{fig:vpython}
\end{center}
\end{figure}

\end{widetext}

The plots in Figures~\ref{fig:colortime} and~\ref{fig:random3bod} are not the clearest depictions of the dynamics. From Eqs.~(\ref{eq:3bodsol1})--(\ref{eq:3bodsol2}) we see that the dynamics should be in general {\it quasiperiodic}. The linear combination of trigonometric functions $a\cos(\omega t) +b\sin(\omega't)$ will only be periodic if the ratio $\frac{\omega}{\omega'}$ is a rational number. One family of periodic solutions is obtained for $\frac{V}{J} = \Delta_{n} = 2\sqrt{3}/n$ for any integer $n$. Figure~\ref{fig:periodic} shows several examples of periodic (closed) orbits. Even in the quasiperiodic case, the instantaneous angular separation between two color charge vectors is always precisely $120^{\circ}$. It is interesting to note that for $n$ divisible by 3, the three trajectories coincide, resulting in a single closed curve along which all three color charges move in the $(Q^{(3)},Q^{(8)})$ plane. Additionally, the point $n=1$ results in the same dynamics as $\Delta =0$. For general values of $\Delta$ not corresponding to periodic motion, the quasiperiodic dynamics leads to space-filling curves that, while being bounded, never close.
The provided \textsc{Jupyter} notebook\cite{supp} also makes use of \textsc{Vpython}\cite{VPython} routines to animate the dynamics with the color charge trajectories traced in real time. While Figure~\ref{fig:periodic} depicts the individual trajectories, the \textsc{VPython} visualization shows the individual color charges with the color assigned according to $(Q^{(3)},Q^{(8)})$ coordinates. Whenever a color charge arrives at an angle $\phi = 30^{\circ}$, it is colored in red with continuous changes in color throughout the evolution resulting from relating $\phi = \tan^{-1}\frac{Q^{(8)}}{Q^{(3)}}$ to the hue angle in a standard HSV color map.\cite{ColorBook} It should be emphasized that the actual dynamics being investigated are of each source's components of color charge. That is, the sources themselves are not moving in real space. To make this more transparent, we fix the three sources as points in space and attach a vector to each source representing its color charge vector. This vector will change magnitude, direction and color in the visualization while the sources remain fixed in place. Several screenshots of the resulting animation for a periodic orbit with $n = 6$ are shown in Figure~\ref{fig:vpython}. In the second panel, one observes a color change process (e.g., what began as the green state is turning into a blue state. The anti-colors shown in Figure~\ref{fig:colorspace} emerge in the third panel.

\section{Discussion}
\label{sec:conclusion}
We have presented a reduced, $(0+1)$-dimensional effective model for the strong interaction in which the components of color charge can be computed as functions of time for a given initial state. Both two- and three-body interactions have been considered, and several choices resulted in compact, analytic solutions. More general types of interactions and initial states have been treated numerically. 

Though the details of the analytic calculations are sometimes fairly tedious, the basic steps involved are no more sophisticated than those used to study spin dynamics in undergraduate quantum mechanics. Accordingly, extensions of the work presented could provide motivation for interesting independent projects. In particular, the spin degree of freedom could also be included to provide a more faithful representation of actual quarks. The attached \textsc{Jupyter} notebook allows one to investigate symmetric three-body interactions, as well as the antisymmetric three-body terms considered here. A much larger phase space is accessible to the curious student than what has been presented.

Another possible line of inquiry is examining how the time required for the off-diagonal charge components to become comparable in magnitude to $q^{(3,8)}$ depends on the relative strength of asymmetric interactions, such as the random couplings considered here. Lastly, investigation of dynamics resulting from the singlet initial configuration is another line of inquiry not developed in this work. The color charge expectation values vanish for all components of all sources at arbitrary time when the initial state is the singlet. That does not imply that nothing interesting happens. One might look to two- or three-point {\it correlation functions} of the charge components for nontrivial dynamics. 

 \begin{acknowledgments}
The authors wish to acknowledge extremely helpful comments and suggestions from the anonymous reviewers which improved the quality of this manuscript significantly. 
\end{acknowledgments}


\begin{thebibliography}{11}
\bibitem{note1} Here ``nonabelian'' refers to the nonabelian (or, noncommutative) nature of the gauge group, $\mbox{SU}(3)$. For two elements $a, b\in \mbox{SU}(3)$, in general one has $a\circ b \neq b\circ a$ where $\circ$ denotes the binary operation of the group. Electrodynamics also possesses a gauge symmetry, but the relevant gauge group is the abelian group $\mbox{U}(1)$ in which elements $a$ may be parameterized by a phase $\phi$ as $a\rightarrow e^{i\phi}$ and the group operation $\circ$ takes the form of ordinary multiplication, which is commutative, or ``abelian.''
\bibitem{note3} Classically, a spin can be thought of as a small loop of current which creates a magnetic field. The interaction of two spins with each other can be pictured as the interaction of one spin with the magnetic field created by the other (or vice versa).
\bibitem{note3a} The actual group elements of \mbox{SU(2)} are written as $g = \exp\left[i\sum_{\alpha}\theta_{\alpha}\sigma^{(\alpha)}\right]$ for some choice of coefficients $\theta_{\alpha}$ with $\alpha = x,y,z$. The requirement that $g$ has determinant equal to unity is equivalent to the requirement that the generators $\sigma^{(\alpha)}$ are {\it traceless}, while unitarity follows from the requirement that the generators are hermitian.
\bibitem{note4} The group $\mbox{SU}(N)$ contains $N^{2}-1$ generators, which gives 3 generators (Pauli matrices) for $\mbox{SU}(2)$ and eight (Gell-Mann matrices) for $\mbox{SU}(3)$.
\bibitem{note6} Note that the equivalent problem in electrostatics is the trivial scenario of two electric charges separated by a fixed distance in which no dynamics occur in any measurable variable.
\bibitem{note7} Python array indices start with 0 instead of 1, and the actual program uses shifted indices so that \texttt{ts[:,:,0]} corresponds to $\hat{t}^{(1)}$, and so on in this way.
\end{thebibliography}
\end{document}